\PassOptionsToPackage{unicode}{hyperref}
\PassOptionsToPackage{hyphens}{url}
\PassOptionsToPackage{dvipsnames,svgnames,x11names}{xcolor}
\documentclass[
  12pt]{article}
\usepackage{amsmath,amssymb}
\usepackage{iftex}
\ifPDFTeX
  \usepackage[T1]{fontenc}
  \usepackage[utf8]{inputenc}
  \usepackage{textcomp} 
\else 
  \usepackage{unicode-math}
  \defaultfontfeatures{Scale=MatchLowercase}
  \defaultfontfeatures[\rmfamily]{Ligatures=TeX,Scale=1}
\fi
\usepackage{lmodern}
\ifPDFTeX\else  
\fi
\IfFileExists{upquote.sty}{\usepackage{upquote}}{}
\IfFileExists{microtype.sty}{
  \usepackage[]{microtype}
  \UseMicrotypeSet[protrusion]{basicmath} 
}{}
\makeatletter
\@ifundefined{KOMAClassName}{
  \IfFileExists{parskip.sty}{%
    \usepackage{parskip}
  }{
    \setlength{\parindent}{0pt}
    \setlength{\parskip}{6pt plus 2pt minus 1pt}}
}{
  \KOMAoptions{parskip=half}}
\makeatother
\usepackage{xcolor}
\makeatletter
\ifx\paragraph\undefined\else
  \let\oldparagraph\paragraph
  \renewcommand{\paragraph}{
    \@ifstar
      \xxxParagraphStar
      \xxxParagraphNoStar
  }
  \newcommand{\xxxParagraphStar}[1]{\oldparagraph*{#1}\mbox{}}
  \newcommand{\xxxParagraphNoStar}[1]{\oldparagraph{#1}\mbox{}}
\fi
\ifx\subparagraph\undefined\else
  \let\oldsubparagraph\subparagraph
  \renewcommand{\subparagraph}{
    \@ifstar
      \xxxSubParagraphStar
      \xxxSubParagraphNoStar
  }
  \newcommand{\xxxSubParagraphStar}[1]{\oldsubparagraph*{#1}\mbox{}}
  \newcommand{\xxxSubParagraphNoStar}[1]{\oldsubparagraph{#1}\mbox{}}
\fi
\makeatother

\usepackage{longtable,booktabs,array}
\usepackage{calc} 
\usepackage{etoolbox}
\makeatletter
\patchcmd\longtable{\par}{\if@noskipsec\mbox{}\fi\par}{}{}
\makeatother
\IfFileExists{footnotehyper.sty}{\usepackage{footnotehyper}}{\usepackage{footnote}}
\makesavenoteenv{longtable}
\usepackage{graphicx}
\makeatletter
\def\maxwidth{\ifdim\Gin@nat@width>\linewidth\linewidth\else\Gin@nat@width\fi}
\def\maxheight{\ifdim\Gin@nat@height>\textheight\textheight\else\Gin@nat@height\fi}
\makeatother
\setkeys{Gin}{width=\maxwidth,height=\maxheight,keepaspectratio}
\makeatletter
\def\fps@figure{htbp}
\makeatother

\makeatletter
\@ifpackageloaded{caption}{}{\usepackage{caption}}
\AtBeginDocument{%
\ifdefined\contentsname
  \renewcommand*\contentsname{Table of contents}
\else
  \newcommand\contentsname{Table of contents}
\fi
\ifdefined\listfigurename
  \renewcommand*\listfigurename{List of Figures}
\else
  \newcommand\listfigurename{List of Figures}
\fi
\ifdefined\listtablename
  \renewcommand*\listtablename{List of Tables}
\else
  \newcommand\listtablename{List of Tables}
\fi
\ifdefined\figurename
  \renewcommand*\figurename{Figure}
\else
  \newcommand\figurename{Figure}
\fi
\ifdefined\tablename
  \renewcommand*\tablename{Table}
\else
  \newcommand\tablename{Table}
\fi
}
\@ifpackageloaded{float}{}{\usepackage{float}}
\floatstyle{ruled}
\@ifundefined{c@chapter}{\newfloat{codelisting}{h}{lop}}{\newfloat{codelisting}{h}{lop}[chapter]}
\floatname{codelisting}{Listing}

\makeatother
\makeatletter
\@ifpackageloaded{caption}{}{\usepackage{caption}}
\@ifpackageloaded{subcaption}{}{\usepackage{subcaption}}
\makeatother

\ifLuaTeX
  \usepackage{selnolig}  
\fi
\usepackage[]{natbib}
\usepackage{bookmark}

\IfFileExists{xurl.sty}{\usepackage{xurl}}{} 
\hypersetup{
  pdftitle={Title},
  pdfauthor={Author 1; Author 2},
  pdfkeywords={3 to 6 keywords, that do not appear in the title},
  colorlinks=true,
  linkcolor={blue},
  filecolor={Maroon},
  citecolor={Blue},
  urlcolor={Blue},
  pdfcreator={LaTeX via pandoc}}

\newcommand{\anon}{1}

\newcommand{\methodname}[1]{\mbox{SVA EIV DAG}}

\usepackage{enumerate}
\usepackage[font=tiny,labelfont=bf]{caption}
\usepackage{tikz}
\usepackage{tikz-network}
\usepackage{pgfplots}
\usepackage{graphicx}
\usepackage{amsmath}
\usepackage{amsbsy}
\usepackage{amssymb}
\usepackage{amsfonts}
\usepackage{mathtools}
\usepackage{booktabs}
 \usepackage{relsize}
\usepackage{array}
\usepackage{comment}
\usepackage{bm}
\usepackage{algpseudocode}

\usepackage[ruled,vlined]{algorithm2e}
\algnewcommand\INPUT{\item[\textbf{Input:}]}%
\algnewcommand\OUTPUT{\item[\textbf{Output:}]}%
\usepackage{xr}
\newcommand{\ind}{\perp\!\!\!\!\perp} 
 
\usepackage{cleveref}
\usepgfplotslibrary{groupplots,dateplot}
\usetikzlibrary{patterns,shapes.arrows}
\usetikzlibrary{arrows.meta}
\pgfplotsset{compat=1.17}
\usepackage{amsthm}
\def\1{{\mathbf 1}}
\def\0{{\mathbf 0}}

\newcommand{\norm}[1]{\left\lVert#1\right\rVert}
\newcommand{\Fnorm}[1]{\left\lVert#1\right\rVert_F}

\newcommand{\twonorm}[1]{\left\lVert#1\right\rVert_2}
\newcommand{\onenorm}[1]{\left\lVert#1\right\rVert_1}

\definecolor{cerulean}{rgb}{0.0, 0.48, 0.65}
\newtheorem{thm}{Theorem}

\newtheorem{cor}{Corollary}

\newtheorem{assumption}{Assumption}

\newtheorem{prop}{Proposition}
\newtheorem{defn}{Definition}

\crefname{figure}{}{}
\crefname{algorithm}{}{}
\Crefname{figure}{}{}
\Crefname{section}{}{}

\newcommand{\ignore}[1]{}

\begin{document}

\def\spacingset#1{\renewcommand{\baselinestretch}%
{#1}\small\normalsize} \spacingset{1}



\if1\anon
{
  \title{\bf 
    Causal Path Analysis from Perturbational and Population-Scale Single-Cell Data with Multiscale Confounding and Measurement Error
    }
  \author{Kwangmoon Park \hspace{.2cm}
    and \hspace{.2cm} Hongzhe Li \thanks{
    Corresponding author}\\
    \hspace{.2cm}
    Department of Biostatistics, Epidemiology, and Informatics,\\ University of Pennsylvania, Philadelphia, PA, USA, 19104
    }
  \maketitle
} \fi

\if0\anon
{
  \bigskip
  \bigskip
  \bigskip
  \begin{center}
    {\LARGE\bf Causal Path Analysis from Perturbational and Population-Scale Single-Cell Data with Multiscale Confounding and Measurement Error}
\end{center}
  \medskip
} \fi

\bigskip
\begin{abstract}
Single-cell perturbation experiments provide causal information on gene regulation, whereas population-scale single-cell studies characterize gene expression and phenotypes in human populations. We develop a framework that integrates these complementary data sources for causal path analysis. Rather than assuming that a perturbational gene network transfers directly to the target population, we use externally learned ancestral relationships to constrain the network topology and re-estimate its direct edges and effects from population data. To address latent heterogeneity and measurement error in multiscale single-cell measurements, we develop a surrogate-variable procedure operating at both the cell and subject levels, combined with errors-in-variables correction for network and outcome regressions. We establish theoretical guarantees for confounder recovery and high-dimensional estimation of network and gene–outcome effects. Simulations demonstrate the importance of jointly correcting confounding and measurement error. An application to acute myeloid leukemia identifies distinct regulatory pathways linking transcriptional regulators to blast count.
\end{abstract}
\noindent%
{\it Keywords: Causal inference,  Population-scale data, Single cell genomics, Structure equation models, Unmeasured confounding.}
\vfill

\newpage
\spacingset{1.8} 

\section{Introduction}

Single-cell perturbation experiments and population-scale single-cell studies provide complementary views of gene regulation. Perturb-seq and related CRISPR-based assays perturb candidate regulators in controlled cellular systems, providing information about the direction of gene-regulatory relationships \citep{dixit2016perturb, replogle2022mapping}. Population-scale single-cell RNA sequencing (scRNA-seq), by contrast, profiles primary cells from many individuals and can link gene expression to subject-level clinical outcomes (or phenotypes) \citep{van2019single}. Integrating these data sources offers an important opportunity: perturbational studies can inform the structure of gene-regulatory networks \citep{barry2021sceptre, brown2025large, park2026causal}, while population data can reveal how those relationships operate in human populations and propagate toward clinically relevant outcomes.

This integration is statistically challenging. Perturb-seq experiments are generally conducted in restricted experimental systems, and their complete regulatory networks and effect sizes may not transfer directly to primary human cells. Population-scale scRNA-seq data are collected in the target population but are observational. Consequently, the gene-regulatory relationships is generally identifiable only up to a Markov equivalence class without additional assumptions or external information \citep{vermapearl, pearl1995causal}. A principled analysis must therefore borrow structural information from perturbational data without assuming that the entire network remains unchanged in the target population.

\begin{figure}[!ht]\centering
   \includegraphics[width=1\textwidth]{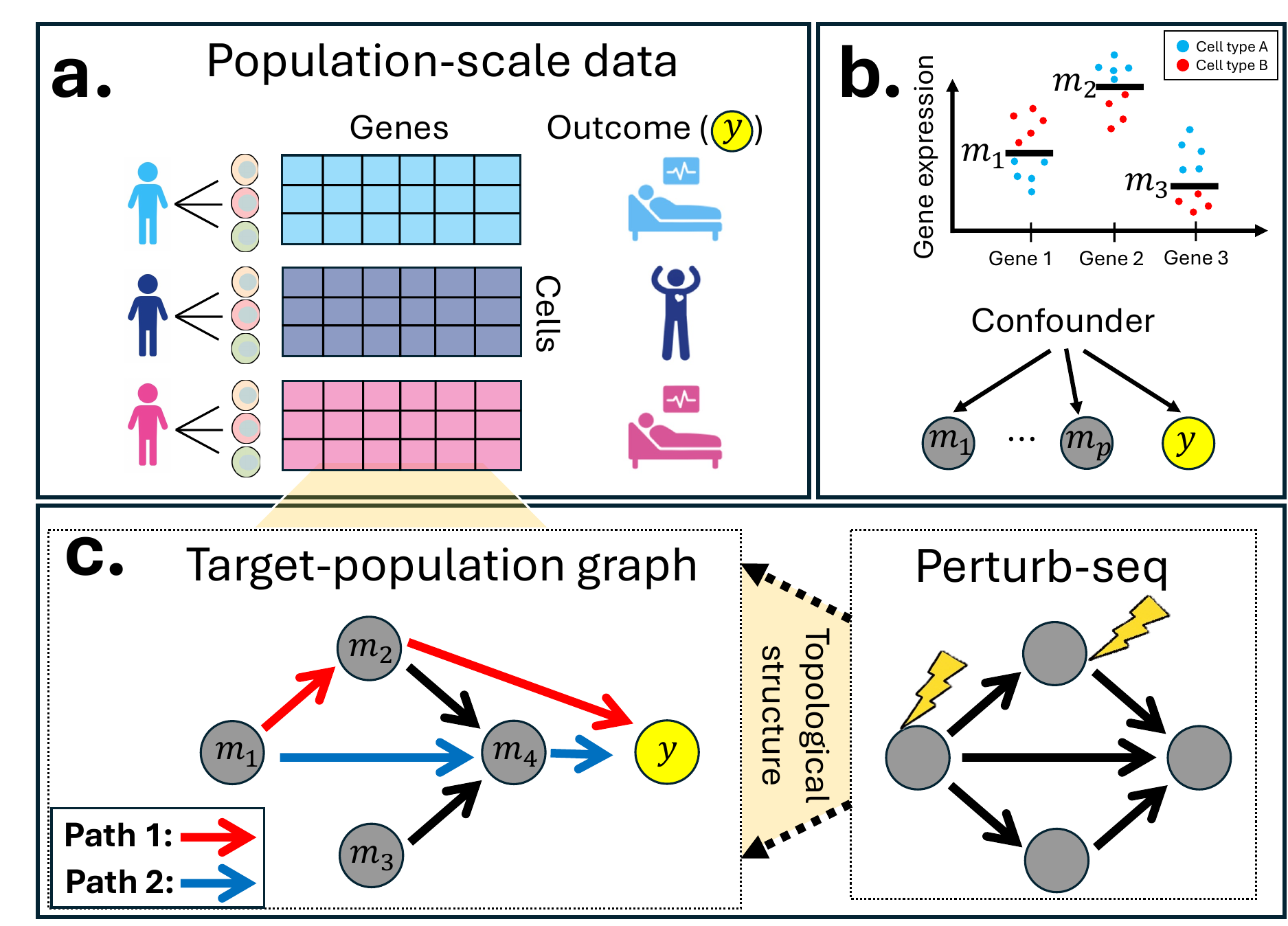}
\caption{\textbf{Schematic overview of the data and statistical problem.} \textbf{a.} Illustration on the population-scale scRNA-seq data. \textbf{b.} Major statistical problems existing in the data: i) An individual's true expression level of genes are measured repeatedly across cells with cell-cell variability stemming from hidden or observed factors, ii) Clinical level confounders can affect both the gene expressions and outcome variable. \textbf{c.} The main task is to estimate the entire graph involving expression levels of genes ($\bm{m}_i$) and the clinical outcome $y_i$ from the population-scale data and by borrowing topological structure of gene network from Perturb-seq experiments.} \label{fig:intro}
\end{figure}

The multiscale structure of population scRNA-seq creates further complications. Gene expression is observed repeatedly at the cellular level, whereas clinical outcomes and many important covariates are measured at the subject level (\textbf{Fig.}~\ref{fig:intro}a). Cellular measurements contain technical noise and are affected by latent cell states, unmeasured cell subtypes, and other sources of within-subject heterogeneity (\textbf{Fig.}~\ref{fig:intro}b). Averaging expression across cells does not eliminate these problems. The resulting subject-level expression summaries remain measured with error, producing errors-in-variables bias when used as predictors \citep{hwang1986multiplicative, carroll1995measurement, xu2007covariate}. Moreover, unmeasured subject characteristics, such as genetic ancestry and environmental exposures, may affect both gene expression and the clinical outcome (\textbf{Fig.}~\ref{fig:intro}b). These subject-level factors can confound the estimated gene network and gene–phenotype relationships.

These challenges are particularly consequential when the scientific target is not merely the association between one gene and an outcome, but the causal pathways through which a regulator may act. A gene may influence a phenotype through several downstream routes, and these pathways may carry effects in different directions. A conventional outcome regression summarizes these mechanisms in a single coefficient and cannot distinguish a potentially actionable pathway from competing or offsetting pathways. Estimating path-specific effects requires recovering the relevant target-population network while addressing measurement error and unmeasured confounding at both the cell and subject levels.

In this paper, we develop a statistical framework for causal path analysis that integrates external perturbational information with confounded, replicated measurements from a target population. Rather than transferring a complete Perturb-seq network, we transfer only an ancestral scaffold that provides candidate upstream–downstream relationships among genes (\textbf{Fig.}~\ref{fig:intro}c). We then use the population-scale data to re-estimate the direct edges and their coefficients (\textbf{Fig.}~\ref{fig:intro}c). This strategy allows regulatory strengths and the presence of direct edges to differ between the experimental system and the target population.

Our approach relies on a structural transportability condition: the external ancestral sets must contain the true parents in the target-population network and preserve their upstream–downstream ordering. Under this condition, the external scaffold resolves the principal orientation ambiguity of observational directed acyclic graph learning while leaving the target-population data responsible for estimating the direct network. 

To address multiscale confounding and measurement error, we propose a surrogate-variable analysis combined with errors-in-variables correction. At the cell level, latent factors are estimated from within-subject measurements to separate genuine cellular heterogeneity from technical noise. The replicated measurements are then used to estimate technical-error variances. At the subject level, latent factors are estimated to account for the unmeasured confounding bias in gene network and outcome regression.
Finally, the estimated confounders and technical-error covariance are incorporated into corrected high-dimensional regressions for the gene network and clinical outcome. This yields direct network effects, total effects of genes on the outcome, and effects transmitted through specific directed paths.

Our work connects several areas of statistical methodology. Existing methods for directed acyclic graph learning recover a graph from observational data only under additional distributional or structural assumptions, or use intervention targets when experimental data are directly available \citep{spirtes2000causation,zheng2018dags,hauser2012characterization}. Our setting differs because the perturbational and target-population data arise from distinct biological systems: the former informs the candidate causal ordering, whereas the latter is used to estimate the target network and its clinical consequences. Classical errors-in-variables methods correct regression bias caused by noisy predictors, but generally consider a single regression with known error covariance and do not address multiscale latent confounding \citep{loh2011high, datta2017cocolasso}. Surrogate-variable methods estimate latent factors, but are typically formulated for a single multivariate outcome model rather than a collection of topology-constrained structural equations with measurement error \citep{leek2007capturing, lee2017improved}. The proposed framework brings these components together for causal path analysis with replicated measurements.

Our main contributions are threefold. First, we formulate a multiscale causal model that integrates external perturbational information with replicated measurements from a target population, using externally learned ancestral relationships to constrain but not determine the target-population DAG. Second, we develop a two-scale surrogate-variable procedure together with errors-in-variables correction for high-dimensional estimation of the network and gene--outcome effects. Third, we establish theoretical guarantees for confounder recovery and causal-effect estimation, and demonstrate the method through simulations and an application to acute myeloid leukemia.

The proposed framework is applicable beyond single-cell genomics. Its essential structure arises whenever noisy, replicated measurements are available in a target population, outcomes are recorded at a higher level of aggregation, and external experiments provide partial information about the causal ordering of the measured variables. Examples may include spatial and multi-omic studies, repeated biomarker measurements, and other settings that combine external perturbational evidence with observational population data.

The paper is organized as follows. Section 2 introduces the statistical model and our target parameters. Section 3 explains how measurement error and unmeasured confounding invalidate standard analyses. Section 4 develops the surrogate-variable procedures for recovering unmeasured confounders at two scales. Section 5 introduces the EIV-corrected network and outcome estimators and establishes their theoretical properties. Sections 6 and 7 report the simulation studies and acute myeloid leukemia application, respectively. Section 8 concludes with limitations and directions for future research.

\subsection{Notation and Preliminaries}

For a positive integer $K$, we write $[K]\coloneqq\{1,\dots,K\}$. Lowercase
boldface letters (e.g. $\bm{a}$) denote vectors, uppercase boldface letters (e.g. $\bm{A}$) denote matrices. For $\bm{a}\in\mathbb{R}^d$, $\onenorm{\bm{a}}=\sum_k|a_k|$, $\norm{\bm{a}}_{\rm max}=\max_k|a_k|$, $\norm{\bm{a}}_0$ counts its nonzero
entries, and $\mathrm{supp}(\bm{a})\coloneqq\{k:a_k\neq0\}$. For a matrix
$\bm{A}$, $\twonorm{\bm{A}}$ denotes the spectral norm, $\norm{\bm{A}}_F$
the Frobenius norm, $\norm{\bm{A}}_{\rm max}=\max_{k,l}|A_{kl}|$ the
entrywise max norm, $\sigma_k(\bm{A})$ its $k$th largest singular value
(with $\sigma_{\max}(\bm{A})\coloneqq\sigma_1(\bm{A})$ and
$\sigma_{\min}(\bm{A})$ its smallest singular value).  
We write $a\vee b=\max(a,b)$, and use $\ind$ for statistical
independence. For a matrix $\bm{A}$ with full column rank, $\bm{P}_{\bm{A}}\coloneqq
\bm{A}(\bm{A}^\top\bm{A})^{-1}\bm{A}^\top$ denotes the orthogonal projection
onto $\mathrm{col}(\bm{A})$. For $\hat{\bm{A}},\bm{A}$ with the same
number of columns, $\|\sin\Theta(\bm{A},\hat{\bm{A}})\|_2\coloneqq
\|\bm{P}_{\bm{A}}-\bm{P}_{\hat{\bm{A}}}\|_2$. For a random variable $X$, $\norm{X}_{\psi_2}$ and $\norm{X}_{\psi_1}$ denote
its sub-Gaussian and sub-exponential norms, respectively
\citep{vershynin2018high}. We write $a_n \gg
b_n$ if $b_n/a_n\to0$, and $a_n\gtrsim b_n$ if $a_n\ge c b_n$ for some
constant $c>0$. The causal dependence structure among the $p$ variables is represented by
a DAG $\mathcal G=(V,E)$ with node set $V=[p]$ and edge
set $E\subseteq V\times V$ without cycles. A directed edge $i\to j$ encodes a direct causal effect of
variable $i$ on variable $j$. For $j\in[p]$, we write $pa(j)$ for its
parents (nodes with an edge into $j$), $anc(j)$ for its ancestors (nodes
with a directed path into $j$), and $des(j)$ for its descendants (nodes
reachable from $j$ by a directed path). 

\section{Problem formulation and identification}

\subsection{Data structure and multiscale measurement model}\label{sec:scmodel}

For subject $i\in[N]$, let
$X^{(i)}\in\mathbb{R}^{n_i\times p}$ denote the single-cell expression measurements from $n_i$ cells across $p$ genes, and let $y_i\in\mathbb{R}$ denote a subject-level outcome. Thus, gene expression is observed at the cell level, whereas the outcome is measured at the subject level.

We model the observed expression of gene $j$ in cell $c$ of subject $i$ of $X^{(i)}$ as
\begin{equation}\label{eq:cell_model}
    x_{ijc} = m_{ij} + \bm{\rho}_j^\top \bm{v}_{ic} + \bm{\alpha}_j^\top \bm{f}_{ic} + \delta_{ijc}\in\mathbb{R},
\end{equation}
where $m_{ij}$ is the true hidden expression level, $\bm{v}_{ic} \in \mathbb{R}^l$ the measured cell-level covariates, $\bm{f}_{ic} \in \mathbb{R}^r$ the unmeasured cell-level factors, and $\delta_{ijc}$ the measurement error.

The latent state $m_{ij}$ represents the subject-level gene expression entering the causal network and outcome model introduced in the following section. Averaging measurements across cells reduces but does not eliminate the effects of latent cellular heterogeneity and technical measurement error; consequently, the resulting subject-level summaries remain subject to confounding and errors-in-variables bias \citep{hwang1986multiplicative,carroll1995measurement,xu2007covariate}.

\subsection{Target-population causal model and causal estimands}\label{sec:targetmodel}
We assume that the latent subject-level expression states follow a linear structural equation model (SEM) associated with a DAG \(\mathcal G\). For gene $j$, with the structural noise $\varepsilon_{ij}$, we have
\begin{equation}\label{eq:local_sem}
    m_{ij} = \mu_j + \bm{\gamma}_j^\top \bm{z}_i + \bm{\lambda}_j^\top \bm{u}_i + \bm{\beta}_{pa(j)}^\top \bm{m}_{i,pa(j)} + \varepsilon_{ij},
\end{equation}
where $\bm{z}_i \in \mathbb{R}^k$ and $\bm{u}_i \in \mathbb{R}^m$ denote measured and unmeasured subject-level confounders, respectively. Here $\mu_j \in
\mathbb{R}$, $\bm{\gamma}_j \in \mathbb{R}^k$, $\bm{\lambda}_j \in \mathbb{R}^m$
, and $\bm{\beta}_{pa(j)}$ is the causal DAG parameter. 

The SEM \eqref{eq:local_sem} across all subjects and genes can be equivalently
written as
\begin{equation}\label{eq:global_sem}
    \bm{M} = \bm{1}\bm{\mu}^\top + \bm{Z}\bm{\Gamma} + \bm{U}\bm{\Lambda} + \bm{M}\bm{B} + \bm{E},
\end{equation}
where $\bm{M} = (m_{ij}) \in
\mathbb{R}^{N \times p}$, $\bm{Z} = (\bm{z}_1^\top, \dots,
\bm{z}_N^\top)^\top \in \mathbb{R}^{N \times k}$, $\bm{U} =
(\bm{u}_1^\top, \dots, \bm{u}_N^\top)^\top \in \mathbb{R}^{N \times
m}$, and $\bm{E} = (\bm{\varepsilon}_1^\top, \dots,
\bm{\varepsilon}_N^\top)^\top \in \mathbb{R}^{N \times p}$. We also
define $\bm{\Gamma} = (\bm{\gamma}_1^\top, \dots,
\bm{\gamma}_p^\top)^\top \in \mathbb{R}^{k \times p}$, $\bm{\Lambda}
= (\bm{\lambda}_1^\top, \dots, \bm{\lambda}_p^\top)^\top \in
\mathbb{R}^{m \times p}$, and $\bm{B} \in \mathbb{R}^{p \times p}$
is the upper-triangular matrix governing the
DAG.

We assume that a subject-level outcome $y_i \in \mathbb{R}$ has a
direct relationship with the latent network variables $\bm{m}_i$
through a linear model with measurement error $\varepsilon_{yi}$:
\begin{equation}\label{eq:outcome_reg}
    y_i = \mu_y + \bm{m}_i^\top \bm{\beta}_m + \bm{z}_i^\top \bm{\beta}_z + \bm{u}_i^\top \bm{\beta}_u + \varepsilon_{yi}.
\end{equation}
Without loss of generality, it suffices to include $\bm{z}_i$ and
$\bm{u}_i$ from \eqref{eq:local_sem}, as other covariates
irrelevant to $\bm{m}_i$ will not bias the estimation of
$\bm{\beta}_m$. We assume no reverse causation between the outcome and gene expression: $y$ does not causally affect any of $m_1,\ldots,m_p$. 

Our main targets of interest in this paper are the causal path effects propagating from the network variables to the outcome. We begin with the total causal effect (TCE).

\begin{defn}[Total causal effect]\label{def:tce_operational}
For $j\in[p]$ and $m\in\mathbb{R}$, let $do(m_j=m)$ denote the intervention that replaces the structural
equation for $m_j$ in \eqref{eq:local_sem} with $m_{ij}\coloneqq m$ \citep{pearl1995causal}.
The total causal effect (TCE) of variable $j$ on the outcome $y$ is
\begin{equation*}
    \mathrm{TCE}_j\ \coloneqq\ \frac{\partial}{\partial m} \mathbb{E}\big[y_i\mid do(m_j=m),\bm{z}_i,\bm{u}_i\big].
\end{equation*}
\end{defn}

We now decompose
$\mathrm{TCE}_j$ into contributions from individual directed paths in $\mathcal G^+$, where $\mathcal G^+$ denote $\mathcal G$ augmented with the outcome edges
$\{j\to y : \beta_{m,j}\neq0\}$. Recall that \eqref{eq:local_sem} can be
written $m_{ij}=\sum_{l\in pa(j)}B_{l,j} m_{i,l}+\kappa_{i,j}$, where
$\kappa_{i,j}\coloneqq\mu_j+\bm\gamma_j^\top\bm z_i+\bm\lambda_j^\top\bm u_i+\varepsilon_{ij}$.
Similarly, \eqref{eq:outcome_reg} can be written
$y_i=\sum_{j\in[p]}\beta_{m,j} m_{ij}+\kappa_{i,y}$, where
$\kappa_{i,y}\coloneqq\mu_y+\bm\beta_z^\top\bm z_i+\bm\beta_u^\top\bm u_i+\varepsilon_{yi}$. 

We first define directed paths in $\mathcal G^+$ and a path-restricted intervention:

\begin{defn}[Directed path]\label{def:path}
A directed path from $j_1$ to $y$ is a sequence of distinct nodes
$\pi=(j_1,j_2,\dots,j_k,y)$ with $B_{j_l,j_{l+1}}\neq0$ for $l\in[k-1]$
and $\beta_{m,j_k}\neq0$. Let $\Pi_{j_1\to y}$ denote the set of all such
paths in $\mathcal G^+$.
\end{defn}

\begin{defn}[$\pi$-restricted intervention]\label{def:pi_do}
For $\pi=(j_1,\dots,j_k,y)\in\Pi_{j_1\to y}$ and $m\in\mathbb R$, define
$m_{i,j_1}\mid do_\pi(m_{j_1}=m)\coloneqq m$ and, recursively for $l=2,\dots,k$,
\begin{align}
m_{i,j_l}\mid do_\pi(m_{j_1}=m) &\coloneqq B_{j_{l-1},j_l} \big(m_{i,j_{l-1}}\mid do_\pi(m_{j_1}=m)\big)
 +\!\!\sum_{r\in pa(j_l)\setminus\{j_{l-1}\}}\!\! B_{r,j_l} m_{i,r}+\kappa_{i,j_l}, \label{eq:pi_do_recursive}\\
y_i \mid do_\pi(m_{j_1}=m) &\coloneqq \beta_{m,j_k} \big(m_{i,j_k}\mid do_\pi(m_{j_1}=m)\big)
 + \sum_{r\neq j_k} \beta_{m,r} m_{i,r} + \kappa_{i,y}. \label{eq:pi_do_outcome}
\end{align}
\end{defn}

With Definitions~\ref{def:path} and~\ref{def:pi_do}, we formally define the
path-specific causal effect (PCE) as follows.

\begin{defn}[Path-specific causal effect]\label{def:pce_operational}
$$\mathrm{PCE}_{\pi} \coloneqq \frac{\partial}{\partial m} 
\mathbb E\big[y_i \mid do_\pi(m_{j_1}=m),\bm{z}_i,\bm{u}_i\big].$$
\end{defn}

The PCE and TCE admit the following closed-form representation, which makes them identifable from the model.

\begin{prop}\label{prop:pce_tce_closed_form}
For every $j\in[p]$ and $\pi=(j_1,j_2,\dots,j_k,y)\in\Pi_{j_1\to y}$ with $j_1=j$,
\begin{equation*}
\mathrm{PCE}_{\pi}=\Big(\prod_{l=1}^{k-1} B_{j_l,j_{l+1}}\Big) \beta_{m,j_k}
\quad \mbox{and} \quad \mathrm{TCE}_j   =   \big\{(\bm{I}-\bm{B})^{-1}\big\}_{[j, :]} \bm{\beta}_m   =   \sum_{\pi \in \Pi_{j\to y}} \mathrm{PCE}_\pi.
\end{equation*}
\end{prop}

\ignore{\begin{prop}\label{prop:pce_closed_form}
For every $\pi=(j_1,j_2,\dots,j_k,y)\in\Pi_{j_1\to y}$,
$$\mathrm{PCE}_{\pi}   =   \Big(\prod_{l=1}^{k-1} B_{j_l,j_{l+1}}\Big) \beta_{m,j_k}.$$
\end{prop}

\begin{proof}
Deferred proof can be found in Section~\ref{supp-proof:pce_closed_form}.
\end{proof}

\begin{prop}[TCE as a sum of PCEs]\label{prop:tce_closed_form}
For every $j\in[p]$,
$$\mathrm{TCE}_j   =   \big\{(\bm{I}-\bm{B})^{-1}\big\}_{[j, :]} \bm{\beta}_m   =   \sum_{\pi \in \Pi_{j\to y}} \mathrm{PCE}_\pi.$$
\end{prop}

\begin{proof}
Deferred proof can be found in Section~\ref{supp-proof:tce_closed_form}.
\end{proof}
}

\subsection{External perturbational information and structural transportability}

Without additional assumptions or information, observational scRNA-seq data identify a DAG up to a Markov equivalence class \citep{vermapearl}. We therefore use external perturbational data, such as Perturb-seq \citep{dixit2016perturb}, to provide information on the topological ordering among genes \citep{brown2025large,park2026causal}. With the externally identified candidate set $\mathcal A_j^{P}$ for gene $j$, we require the structural transportability condition
\begin{equation*}
\operatorname{pa}(j)\subseteq \mathcal A_j^{P}\subseteq [p]\setminus(\{j\}\cup \operatorname{des}(j)), \qquad j\in[p].    
\end{equation*}
The lower bound ensures no true parent is missed; the upper bound excludes descendants. Throughout, we let $\mathcal A_j^P=anc(j)$ without loss of generality, since every subsequent derivation relies on the condition and goes through unchanged with $\mathcal A_j^P$ in place of $anc(j)$. The target-population data then select the direct edges from $anc(j)$ and re-estimate their coefficients.

\section{Errors-in-variables bias and confounding}
\label{sec:stat_challenge}

 \subsection{Errors-in-variables bias in estimating the causal effects}

Estimating the causal estimands requires estimating $\bm{B}$ and
$\bm{\beta}_m$ from subject-level data. Nonetheless, the estimation is challenging in several aspects. We clarify that the regressors are corrupted in subject-level SEM regression and outcome regressions.

First, averaging the cell-level observation model over
the $n_i$ cells of subject $i$ yields:
\begin{equation}\label{eq:subject_mean}
    \bar{x}_{ij} = m_{ij} + \bm{\rho}_j^\top \bar{\bm{v}}_i + \bm{\alpha}_j^\top \bar{\bm{f}}_i + \bar{\delta}_{ij},
\end{equation}
where $\bar{\bm{v}}_i = \frac{1}{n_i}\sum_c \bm{v}_{ic}$,
$\bar{\bm{f}}_i = \frac{1}{n_i}\sum_c \bm{f}_{ic}$, and
$\bar{\delta}_{ij} = \frac{1}{n_i}\sum_c \delta_{ijc}$. We then define the latent noiseless biological state as 
$\bm{x}^*_i = \bm{m}_i + \bm{P}^\top\bar{\bm{v}}_i + \bm{A}^\top\bar{\bm{f}}_i \in \mathbb{R}^p,$
where $\bm{P}=(\bm{\rho}_1^\top,\cdots,\bm{\rho}_p^\top)^\top\in \mathbb{R}^{p\times l}$,
$\bm{A}=(\bm{\alpha}_1^\top,\cdots,\bm{\alpha}_p^\top)^\top\in \mathbb{R}^{p\times r}$. Plugging the SEM \eqref{eq:local_sem} into \eqref{eq:subject_mean} yields:
\begin{align}
\bar{x}_{ij}
    = \mu + \bm{\gamma}_{aug}^{\top} \bm{z}_{i,aug}
      + \bm{\lambda}_{aug}^{\top} \bm{u}_{i,aug}
      + \bm{\beta}_{pa(j)}^\top \bm{x}^*_{i,pa(j)}
      + \varepsilon_{ij} + \bar{\delta}_{ij}, \label{eq:subject_structural}
\end{align}
where $\bm{z}_{i,aug}= [\bm{z}_i^\top, \bar{\bm{v}}_i^\top]^\top$,
$\bm{u}_{i,aug} = [\bm{u}_i^\top, \bar{\bm{f}}_i^\top]^\top$, and
$\bm{\gamma}_{aug}= [\bm{\gamma}_j^\top,\ (\bm{\rho}_j -
\bm{P}_{pa(j)}^\top \bm{\beta}_{pa(j)})^\top]^\top$ and $\bm{\lambda}_{aug}= [\bm{\lambda}_j^\top,\ (\bm{\alpha}_j -
\bm{A}_{pa(j)}^\top \bm{\beta}_{pa(j)})^\top]^\top$ and $\bm{x}^*_{i,pa(j)}\in\mathbb{R}^{|pa(j)|}$ is a $pa(j)$ entry extraction of $\bm{x}_i^*$.

Likewise, the outcome regression \eqref{eq:outcome_reg} can be formulated as:
\begin{align}
    y_i 
        = \mu_y+\bm{x}^{*\top}_i \bm{\beta}_m + \bm{z}_{i,aug}^\top \bm{\beta}_{z,aug} + \bm{u}_{i,aug}^\top \bm{\beta}_{u,aug} + \varepsilon_{yi}, \label{eq:outcome_eiv_full}
\end{align}
where $\bm{\beta}_{z,aug} = [\bm{\beta}_z^\top, -(\bm{P}^\top\bm{\beta}_m)^\top]^\top$
and $\bm{\beta}_{u,aug} = [\bm{\beta}_u^\top, -(\bm{A}^\top\bm{\beta}_m)^\top]^\top$.

\ignore{Since $\bar{\bm{x}}_i = \bm{x}^*_i + \bar{\bm{\delta}}_i$, the OLS estimator 
$\hat{\bm{\beta}}_m^{\mathrm{OLS}}\coloneqq \left(\sum_{i=1}^N\bar{\bm{x}}_{i}\bar{\bm{x}}_{i}^\top\right)^{-1}
\sum_{i=1}^N\bar{\bm{x}}_{i}y_i$ becomes
inconsistent:
\begin{equation}\label{eq:outcome_attenuation}
\twonorm{\hat{\bm{\beta}}_m^{\mathrm{OLS}}-\left(\bm{\Sigma}_{\bm{x}^*} + \bar{\bm{\Sigma}}_\delta\right)^{-1}\bm{\Sigma}_{\bm{x}^*}\bm{\beta}_m}=o_p(1),
\end{equation}
where $\bar{\bm{\Sigma}}_\delta = \frac{1}{N}\sum_{i=1}^N \frac{1}{n_i}\bm{\Sigma}_{\delta_i}$.}

In both the SEM regression \eqref{eq:subject_structural} and the outcome regression \eqref{eq:outcome_eiv_full}, the noiseless regressors $\bm{x}^*_{i,pa(j)}$ and $\bm{x}^*_{i}$ remain unobserved. Instead, we observe the noisy subject-specific means
\begin{align*}
\bar{\bm{x}}_{i,pa(j)}= \bm{x}^*_{i,pa(j)} + \bar{\bm{\delta}}_{i,pa(j)}\quad\mbox{and}\quad
\bar{\bm{x}}_{i}= \bm{x}^*_{i} + \bar{\bm{\delta}}_{i},
\end{align*}
respectively, where $\bar{\bm{\delta}}_{i} = \frac{1}{n_i}\sum_c \bm{\delta}_{ic}$, with $\bm{\delta}_{ic} \in \mathbb{R}^{p}$ the vector concatenating $\delta_{ikc}$ across $k \in [p]$, and $\bar{\bm{\delta}}_{i,pa(j)}$ denotes the $pa(j)$-entry extraction of $\bar{\bm{\delta}}_{i}$.
\ignore{In model~\eqref{eq:subject_structural}, even when the true parent set $pa(j)$ is known, the noiseless state
$\bm{x}^*_{i,pa(j)}$ remains unobserved. Instead, we observe the
corrupted subject-specific mean $$\bar{\bm{x}}_{i,pa(j)} =
\bm{x}^*_{i,pa(j)} + \bar{\bm{\delta}}_{i,pa(j)},$$ where
$\bar{\bm{\delta}}_{i,pa(j)} = \frac{1}{n_i}\sum_c
\bm{\delta}_{i,pa(j),c}$ and $\bm{\delta}_{i,pa(j),c} \in
\mathbb{R}^{|pa(j)|}$ is the vector concatenating $\delta_{ikc}$ across $k \in
pa(j)$.}
With such errors in the regressors, the estimation of $\bm{\beta}_{pa(j)}$ and $\bm{\beta}_{m}$ remains biased, a phenomenon well documented in the errors-in-variables (EIV) literature \citep{hwang1986multiplicative,carroll1995measurement,xu2007covariate,loh2011high,datta2017cocolasso}. Correcting this bias requires estimating $\mathrm{Cov}(\bm{\delta}_{ic})$. However, the within-subject variation in $x_{ijc}$ reflects both technical error $\delta_{ijc}$ and variation driven by the latent $\bm{f}_{ic}$, motivating the cell-level confounder adjustment in Section~\ref{sec:method}.

\subsection{Unmeasured confounding across scales}

Equations~\eqref{eq:subject_structural} and \eqref{eq:outcome_eiv_full} contain unmeasured variation operating at two distinct scales. First, at the subject scale, the augmented confounder $\bm{u}_{i,\mathrm{aug}}=(\bm{u}_i^\top,\bar{\bm{f}}_i^\top)^\top$ affects both the gene-expression network \eqref{eq:subject_structural} and the outcome \eqref{eq:outcome_eiv_full}. Omitting it from the two regressions induces omitted-variable bias in the estimated coefficients $\bm{B}$ and $\bm{\beta}_m$.

At the cell scale, the cell-level hidden factors $\bm{f}_{ic}$ also affect the estimation of the technical-error covariance. Their within-subject variation contributes to the residual variability
of $x_{ijc}$ and must therefore be separated from the technical error $\delta_{ijc}$; otherwise, genuine biological heterogeneity may be incorrectly attributed to measurement error in the EIV correction.



\section{Surrogate Variable Analysis for Confounders}\label{sec:method}

\subsection{Assumptions for two-scale confounder recovery}

We first state assumptions on the technical and structural errors and on the separation between latent factors and observed covariates, used for two-scale confounder recovery.

\begin{assumption}[Technical and structural errors]\label{assump:noise}
Assume the following for $\bm{\delta}_{ic}$ and $\bm{\varepsilon}_i$:
\begin{itemize}
\vspace{-0.3em}
\item [(i)] (Zero mean, independence): $\mathbb{E}[\bm{\delta}_{ic}]=\bm{0}$, $\mathbb{E}[\varepsilon_{ij}]=0$, $\mathbb{E}[\varepsilon_{yi}]=0$; $\bm{\delta}_{ic}\ind \bm{\delta}_{ic'}$, 
$\bm{\delta}_{ic}\ind \bm{\delta}_{i'c}$, $\varepsilon_{ij}\ind\varepsilon_{i'j}$ and $\varepsilon_{yi}\ind\varepsilon_{yi'}$ for $c\neq c'$ and $i\neq i'$
\item [(ii)] (Feature-wise independence): $\delta_{ijc}\ind \delta_{ij'c}$
$\varepsilon_{ij}\ind \varepsilon_{ij'}$ for $j\neq j'$ 
with diagonal $\mathrm{Cov}(\bm{\delta}_{ic})=\bm{\Sigma}_{\delta_i}$
and $\mathrm{Cov}(\bm{\varepsilon}_i)=\mathrm{diag}(\sigma_1^2,\dots,\sigma_p^2)$.
\item [(iii)] (Exogeneity): $\epsilon_{ij}$, $\bm{\delta}_{ic}$ and $\varepsilon_{yi}$ are mutually independent, $\varepsilon_{ij}\ind (\{m_{ik}\}_{k\notin des(j)},\bm{z}_i,\bm{u}_i,\bm{v}_{ic},\bm{f}_{ic})$ and 
$(\bm{\delta}_{ic},\varepsilon_{yi})\ind (\bm{m}_i,\bm{z}_i,\bm{u}_i,\bm{v}_{ic},\bm{f}_{ic})$ for $\forall c\in [n_i],i\in[N],j\in[p]$.
\item [(iv)] (Sub-Gaussian tails): With $\sigma_\varepsilon^{(0)}<\infty$ and $\sigma_{\max}<\infty$, assume 
\begin{equation}\label{eq:subgaussian}
\sup_{i\in[N]}\sup_{\bm u\in\mathbb{S}^{p-1}}\norm{\langle\bm u,\bm\varepsilon_i\rangle}_{\psi_2}\le \sigma_\varepsilon^{(0)},
\qquad
\sup_{i\in[N],c\in[n_i]}\sup_{\bm u\in\mathbb{S}^{p-1}}\norm{\langle\bm u,\bm\delta_{ic}\rangle}_{\psi_2}\le \sigma_{\max}.
\end{equation}
\end{itemize}
\end{assumption}


\begin{assumption}[Confounder--covariate separation]\label{assump:conf_cov}
Let $\tilde{\bm{U}}_{aug}\coloneqq(\bm{I}_N-\bm{P}_{[\bm{1},\bm{Z}_{aug}]})\bm{U}_{aug}$ and $\tilde{\bm{F}}_{i}\coloneqq(\bm{I}_{n_i}-\bm{P}_{[\bm{1}_{n_i},\bm{V}_i]})\bm{F}_i$ for each $i\in[N]$. The latent factors are not collinear with the measured covariates, equivalently the following full-rank conditions hold after residualization.
\begin{itemize}
\vspace{-0.3em}
\item [(i)] (Subject scale): $\mathrm{rank}\big(\tilde{\bm{U}}_{aug}\big)=m+r$.
\item [(ii)] (Cell scale): $\mathrm{rank}\big(\tilde{\bm{F}}_{i,c}\big)=r$ for every $i\in[N]$.
\end{itemize}
\end{assumption}

This condition requires that the latent factors retain variation after projecting out the observed covariates; otherwise their effects cannot be separately recovered.

\subsection{Subject-scale SVA}\label{sec:subject_sva}
To clarify the impact of unmeasured confounding, we formulate the structural equation. By extending the parent set to the known ancestor set $anc(j)$ with $|anc(j)|=q_j$ obtained from external source, the local SEM in \eqref{eq:subject_structural} for a gene $j$ across all $N$ subjects can be written as:
\begin{equation}\label{eq:anc_regression_xstar}
    \bar{\bm{x}}_j = \bm{1} \mu_j + \bm{Z}_{aug} \bm{\gamma}_{j, aug} + \bm{X}^*_{anc(j)} \bm{\beta}_{anc(j)} + \bm{U}_{aug} \bm{\lambda}_{j, aug} + \bm{\varepsilon}_j + \bar{\bm{\delta}}_j
\end{equation}
where $\bm{Z}_{aug} = [\bm{Z}, \bar{\bm{V}}] \in \mathbb{R}^{N \times (k+l)}$ is the augmented known covariate matrix, $\bm{U}_{aug} = [\bm{U}, \bar{\bm{F}}] \in \mathbb{R}^{N \times (m+r)}$ the augmented unmeasured confounder matrix, and $\bm{X}^*_{anc(j)} \in \mathbb{R}^{N \times q_j}$ the noiseless aggregated ancestor state. The coefficient $\bm{\beta}_{anc(j)}$ represents the link between $j$-th gene and its ancestor nodes, whose non-parent entries are zero. 

Even if we temporarily treat $\bm{X}^*_{anc(j)}$ as perfectly observed, attempting to estimate $\bm{\beta}_{anc(j)}$ directly via standard regression fails, as $\bm{U}_{aug}$ confounds the ancestors and the target gene, inducing omitted variable bias. However, because the confounders operate in a lower-dimensional space with the SEM \eqref{eq:global_sem}, projecting out the observed covariates yields empirical residuals that are dominated by this low-rank confounding subspace. This motivates our suggested Surrogate Variable Analysis (SVA) strategy.

We form a design matrix $\bm{O} = [\bm{1}, \bm{Z}_{aug}]$ and  the orthogonal projection operator
$\bm{P}_{\bm{O}} = \bm{O}(\bm{O}^\top \bm{O})^{-1}\bm{O}^\top$. We then define 
\begin{equation}\label{eq:R_def}
    \bm{R} = (\bm{I}-\bm{P}_{\bm{O}})\bar{\bm{X}} \in \mathbb{R}^{N \times p}.
\end{equation}
which retains
a low-rank signal plus noise structure, where the left signal column space corresponds to the
residualized $\bm{U}_{aug}$. We propose to estimate
$\tilde{\bm{U}}_{aug} = (\bm{I} - \bm{P}_{[\bm{1}, \bm{Z}_{aug}]})\bm{U}_{aug}$ based on rank $m+r$
truncated singular value decomposition (SVD) on $\bm{R}$:
\begin{equation*}
\hat{\bm{U}}_{aug}\leftarrow \mathrm{SVD}_{m+r}(\bm{R}),
\end{equation*}
where $\mathrm{SVD}_{m+r}(\bm{R})$ returns the left $m+r$ singular vector matrix of $\bm{R}$.
Under additional assumptions, we can consistently estimate
the column space of $\tilde{\bm{U}}_{aug}$ as $N,p\rightarrow \infty$.

\begin{thm}[Residual factorization]\label{thm:eiv_factorization}
Under the observation model
\eqref{eq:cell_model}, the structural equation model \eqref{eq:global_sem}, and Assumption~\ref{assump:noise}, assuming the composite confounder
direct effects satisfy $\mathrm{rank}(\bm{\Lambda}_{aug})=m+r$, the residual
matrix $\bm{R}$ defined in \eqref{eq:R_def} satisfies
\begin{equation*}
    \bm{R} = \tilde{\bm{U}}_{aug} \bm{\Lambda}_{aug}(\bm{I}-\bm{B})^{-1} + \tilde{\bm{E}}(\bm{I}-\bm{B})^{-1} + \tilde{\bm{\Delta}},
\end{equation*}
where $\tilde{\bm{U}}_{aug}=(\bm{I}-\bm{P}_{[\bm{1},\bm{Z}_{aug}]})\bm{U}_{aug}$,
$\tilde{\bm{E}}=(\bm{I}-\bm{P}_{[\bm{1},\bm{Z}_{aug}]})\bm{E}$, and
$\tilde{\bm{\Delta}}=(\bm{I}-\bm{P}_{[\bm{1},\bm{Z}_{aug}]})\bar{\bm{\Delta}}$. Here $(\bm{I}-\bm{B})^{-1}$
is an invertible loading matrix, so $\mathrm{rank}\big(\bm{\Lambda}_{aug}(\bm{I}-\bm{B})^{-1}\big)=m+r$.
\end{thm}

\begin{thm}[Asymptotic recovery of the confounder subspace]\label{thm:u_recovery}
Under the models \eqref{eq:cell_model}, \eqref{eq:global_sem}, and under Assumption~\ref{assump:noise} and Assumption~\ref{assump:conf_cov}(i), suppose $\twonorm{(\bm{I}-\bm{B})^{-1}}\le C_B$. If the signal strength of the latent
confounding $\bm{\Phi}\coloneqq\tilde{\bm{U}}_{aug}\bm{\Lambda}_{aug}(\bm{I}-\bm{B})^{-1}$ satisfies
\begin{equation}\label{eq:subject_recovery_hyp}
    \sigma_{m+r}(\bm{\Phi})\gg\sqrt{N}+\sqrt{p},
\end{equation}
then as $N,p\to\infty$, the column space of $\hat{\bm{U}}_{aug}$ converges to the true subspace
$\tilde{\bm{U}}_{aug}$:
\begin{equation*}
    \|\sin\Theta(\tilde{\bm{U}}_{aug},\hat{\bm{U}}_{aug})\|_2 \stackrel{p}{\to} 0.
\end{equation*}
\end{thm}

The theorem demonstrates that $\mathrm{col}(\tilde{\bm{U}}_{aug})$ can be consistently estimated
whenever the latent confounding is pervasive, in the sense that each confounder influences a
non-vanishing fraction of both the $N$ subjects and the $p$ genes. As an example, when
$\sigma_{m+r}(\tilde{\bm{U}}_{aug})\asymp\sqrt{N}$ and $\sigma_{m+r}(\bm{\Lambda}_{aug})\asymp\sqrt{p}$, under $\sigma_{\min}((\bm{I}-\bm{B})^{-1})\ge c_\Theta>0$, we can easily satisfy \eqref{eq:subject_recovery_hyp}.


Then, the $\bm{B}$ and $\bm{\beta}_m$ are identifiable given that we have full access to $\bm{x}_i^*$. Nevertheless, $\bm{x}_{ij}^*$ is corrupted by random error and an appropriate correction requires cell-scale SVA.

\subsection{Subject-specific cell-scale SVA}\label{sec:cell_sva}
The subject-scale SVA recovers the confounder relevant to the network and outcome regressions, but does not separate within-subject heterogeneity $\bm{f}_{ic}$ from technical error \(\bm{\delta}_{ic}\). To correct this, we perform a subject-specific SVA at the cell level. For each subject $i\in [N]$, $\bm{X}^{(i)} \in \mathbb{R}^{n_i \times p}$ can be expressed as:
\begin{equation}
    \bm{X}^{(i)} = \bm{1}_{n_i} \bm{m}_i^{\top} + \bm{V}_i \bm{P}^\top + \bm{F}_i \bm{A}^\top + \bm{\Delta}_i.
\end{equation}
We construct the design matrix $\bm{O}^{(i)} = [\bm{1}_{n_i}, \bm{V}_i]$. Applying the orthogonal projection matrix $\bm{P}_{\bm{O}^{(i)}} = \bm{O}^{(i)}(\bm{O}^{(i)\top} \bm{O}^{(i)})^{-1}\bm{O}^{(i)\top}$ yields the residual matrix $\bm{R}^{(i)} = (\bm{I}_{n_i} - \bm{P}_{\bm{O}^{(i)}})\bm{X}^{(i)}$. We then propose to estimate $\tilde{\bm{F}}_{i} = (\bm{I}_{n_i} - \bm{P}_{\bm{O}^{(i)}})\bm{F}_i$ from a rank $r$ truncated SVD on $\bm{R}^{(i)}$:
\begin{equation*}
\hat{\bm{F}}_{i}\leftarrow \mathrm{SVD}_{r}(\bm{R}^{(i)}),
\end{equation*}
for each $i\in [N]$. Then, as in subject-scale SVA, we can identify the column space of $\tilde{\bm{F}}_{i}$ with additional assumptions on the within-subject confounding effect.
Assuming $\text{rank}(\bm{A}^\top) = r$, the empirical residual matrix $\bm{R}^{(i)} \in \mathbb{R}^{n_i \times p}$ admits the exact factorization:
\begin{equation*}
    \bm{R}^{(i)} = \tilde{\bm{F}}_{i} \bm{A}^\top + \tilde{\bm{\Delta}}_{i}
\end{equation*}
where $\tilde{\bm{F}}_{i} = (\bm{I}_{n_i} - \bm{P}_{\bm{O}^{(i)}})\bm{F}_i$ and $\tilde{\bm{\Delta}}_{i} = (\bm{I}_{n_i} - \bm{P}_{\bm{O}^{(i)}})\bm{\Delta}_i$. 

\begin{thm}[Asymptotic recovery of the cell-level confounder subspace]\label{thm:cell_recovery}
Under the models \eqref{eq:cell_model}, \eqref{eq:global_sem}, and under Assumption~\ref{assump:noise},\ref{assump:conf_cov}(ii), if the signal strength of the latent cell-level confounding ($\bm{\Phi}_i\coloneqq \tilde{\bm{F}}_{i}\bm{A}^\top$) satisfies
\begin{equation}\label{eq:cell_recovery_hyp}
    \sigma_r(\bm{\Phi}_i)\gg \sqrt{n_i}+\sqrt{p},
\end{equation}
then as $n_i, p \to \infty$, the column space of $\hat{\bm{F}}_i$ converges to the true centered subspace $\tilde{\bm{F}}_{i}$:
\begin{equation*}
    \|\sin \Theta(\tilde{\bm{F}}_{i}, \hat{\bm{F}}_i)\|_2 \stackrel{p}{\to} 0.
\end{equation*}
\end{thm}
The cell-scale SVA removes latent within-subject heterogeneity, isolating the technical-error covariance $\bm{\Sigma}_{\delta_i}$ needed for the EIV correction (Section~\ref{EIV.sec}). See
Algorithm~\ref{alg:sva_correction} for implementation and Section~\ref{supp-sec:implementation} for details on rank selection.

\begin{algorithm}[!ht]
\caption{Subject- and Cell-Scale SVA Confounder Recovery}\label{alg:sva_correction}
\textbf{Input:} $\{\bm{X}^{(i)}\}_{i=1}^N$, $\{\bm{V}_i\}_{i=1}^N$ and $\bm{Z}$, ranks $r$ and $m+r$.\\
\textbf{Output:} $\hat{\bm{U}}_{aug}$, $\{\hat{\bm{F}}_i\}_{i=1}^N$, $\bar{\bm{X}}$, $\bm{Z}_{aug}$.

\underline{Step 1 (Aggregation).} Compute $\bar{\bm{x}}_i=\frac{1}{n_i}\sum_{c=1}^{n_i}\bm{x}_{ic}$ and $\bar{\bm{v}}_i=\frac{1}{n_i}\sum_{c}\bm{v}_{ic}$ for $i\in[N]$; set $\bar{\bm{X}}=(\bar{\bm{x}}_1,\ldots,\bar{\bm{x}}_N)^\top$ and $\bm{Z}_{aug}=[\bm{Z},\bar{\bm{V}}]$.

\underline{Step 2 (Cell-scale SVA).} For each $i\in[N]$, form $\bm{O}^{(i)}=[\bm{1}_{n_i},\bm{V}_i]$, compute the residual $\bm{R}^{(i)}=(\bm{I}_{n_i}-\bm{P}_{\bm{O}^{(i)}})\bm{X}^{(i)}$, and set $$\hat{\bm{F}}_i \leftarrow \mathrm{SVD}_r(\bm{R}^{(i)}).$$

\underline{Step 3 (Subject-scale SVA).} For each $j\in[p]$, form $\bm{O}=[\bm{1},\bm{Z}_{aug}]$ and define $\bm{R}'$ based on \eqref{eq:R_def} and set 
$$\hat{\bm{U}}_{aug}\leftarrow \mathrm{SVD}_{m+r}(\bm{R}).$$

\textbf{Return} $\hat{\bm{U}}_{aug}$, $\{\hat{\bm{F}}_i\}_{i=1}^N$, $\bar{\bm{X}}$, $\bm{Z}_{aug}$.
\end{algorithm}

\section{EIV bias correction and causal effect estimation} \label{EIV.sec}

\subsection{EIV formulation and asymptotic bias}\label{sec:bias}
Notice from \eqref{eq:subject_structural} that we have
\begin{align}
    \bar{x}_{ij} &= \mu + \bm{\gamma}_{aug}^{\top} \bm{z}_{i,aug} + \bm{\lambda}_{aug}^{\top} \bm{u}_{i,aug} + \bm{\beta}_{anc(j)}^\top \bm{x}^*_{i,anc(j)} + \varepsilon_{ij} + \bar{\delta}_{ij} \label{eq:shifted_sem},
\end{align}
where $\bm{x}^*_{i,anc(j)}$ is defined as $anc(j)$ entries of $\bm{x}^*_{i}$. We absorb the sparsity of the non-parent entries into $\bm{\beta}_{anc(j)}$, and define $\bm{\gamma}_{aug} \coloneqq (\bm{\gamma}_j - \bm{P}_{anc(j)}^\top \bm{\beta}_{anc(j)})$ and $\bm{\lambda}_{aug} \coloneqq (\bm{\lambda}_j - \bm{A}_{anc(j)}^\top \bm{\beta}_{anc(j)})$.

For simplicitiy, we first partial out the observed and latent nuisance variables, by defining the oracle predictor matrix $\bm{H}^*=[\bm{1},\bm{Z}_{aug},\bm{U}_{aug}^*]\in \mathbb{R}^{N \times d_H}$, where $d_H \coloneqq 1+k+l+m+r$. We define $\bm{U}_{aug}^*=\tilde{\bm{U}}_{aug}\bm{W}$ for an orthogonal rotation matrix $\bm{W} \in \mathbb{R}^{(m+r) \times (m+r)}$, which $\hat{\bm{U}}_{aug}$ converges to by Theorem~\ref{thm:u_recovery}. We also define $\bm{Q}^* = \bm{I}_N - \bm{P}^*$, where
$\bm{P}^* \coloneqq \bm{H}^*(\bm{H}^{*\top}\bm{H}^*)^{-1}\bm{H}^{*\top}$.

Notice that we have $\mathrm{col}(\bm{P}^*)=\mathrm{col}\big([\bm{1},\bm{Z}_{aug},\bm{U}_{aug}]\big)$ and equation \eqref{eq:shifted_sem} can be simplified as:
\begin{align}\label{eq:shifted_sem_fwl}
   \bm{Q}^* \bar{\bm{x}}_{j} = \underbrace{\bm{Q}^*\bm{X}_{anc(j)}^*}_{\mathrm{Oracle\,regressor}}\bm{\beta}_{anc(j)}  + \bm{Q}^*\bm{\varepsilon}_{j} + \bm{Q}^*\bar{\bm{\delta}}_{j} 
\end{align}
across individuals, where $\bm{X}_{anc(j)}^*=(\bm{x}_{1,anc(j)}^{*\top},\cdots,\bm{x}_{N,anc(j)}^{*\top})^\top\in\mathbb{R}^{N\times q_j}$.

In practice, we observe the corrupted ancestor observations $\bar{\bm{X}}_{anc(j)}=\bm{X}_{anc(j)}^*+\bar{\bm{\Delta}}_{anc(j)}$ and the empirical projection matrix $\hat{\bm{Q}}= \bm{I}_N - \hat{\bm{H}}(\hat{\bm{H}}^\top\hat{\bm{H}})^{-1}\hat{\bm{H}}^\top$, where $ \hat{\bm{H}}=[\bm{1},\bm{Z}_{aug},\hat{\bm{U}}_{aug}]$. Then, instead of the oracle predictor $\bm{Q}^*\bm{X}_{anc(j)}^*$ in  \eqref{eq:shifted_sem_fwl}, we have a corrupted predictor:
\begin{align}
    \hat{\bm{Q}} \bar{\bm{X}}_{anc(j)} &= \bm{Q}^* \bm{X}^*_{anc(j)} + (\hat{\bm{Q}} - \bm{Q}^*) \bar{\bm{X}}_{anc(j)} + \bm{Q}^* \bar{\bm{\Delta}}_{anc(j)} \notag \\
    &= \underbrace{\bm{Q}^* \bm{X}^*_{anc(j)}}_{\mathrm{Oracle\,regressor}} + \underbrace{\bm{E}_Q \bar{\bm{X}}_{anc(j)}}_{\mathrm{SVA\,error}} + \underbrace{\bm{Q}^* \bar{\bm{\Delta}}_{anc(j)}}_{\mathrm{Measurement\,error}}.\label{eq:SEM_error}
\end{align}
Here, $\bm{E}_Q\coloneqq \hat{\bm{Q}} - \bm{Q}^*$ denotes SVA error, which we additionally need to control on top of the measurement error $\bar{\bm{\Delta}}_{anc(j)}$.

Similarly, for the outcome regression, the EIV model can be parameterized as:
\begin{equation}\label{eq:shifted_outcome_fwl}
\bm{Q}^* \bm{y} = \underbrace{\bm{Q}^*\bm{X}^*}_{\mathrm{Oracle\,regressor}}\bm{\beta}_m + \bm{Q}^*\bm{\varepsilon}_{y},
\end{equation}
where the ideal predictor $\bm{X}^*=(\bm{x}^{*\top}_1,\cdots,\bm{x}^{*\top}_N)^\top\in\mathbb{R}^{N\times p}$. Likewise, instead of $\bm{Q}^*\bm{X}^*$ in \eqref{eq:shifted_outcome_fwl}, we have the corrupted predictor $\hat{\bm{Q}} \bar{\bm{X}}=\hat{\bm{Q}}(\bm{X}^*+\bar{\bm{\Delta}})$ that can be decomposes as:
\begin{equation}
    \hat{\bm{Q}} \bar{\bm{X}} = \underbrace{\bm{Q}^* \bm{X}^*}_{\mathrm{Oracle\,regressor}} + \underbrace{\bm{E}_Q \bar{\bm{X}}}_{\mathrm{SVA\,error}} + \underbrace{\bm{Q}^* \bar{\bm{\Delta}}}_{\mathrm{Measurement\,error}}\label{eq:outcome_error}.
\end{equation}

The errors contaminating the observed predictors in both the SEM \eqref{eq:SEM_error} and outcome regressions \eqref{eq:outcome_error} bias the estimation of $\bm{B}$ and $\bm{\beta}_m$, respectively. We specify the asymptotic bias induced by EIV to motivate the bias-correction approach we propose in a later section. To guarantee the concentration of the sample moments for the unobserved latent states, we formalize the tail behavior of their underlying distributions.

\begin{assumption}[Sub-Gaussian Nuisance Factors]\label{assump:subgaussian}
The subject-level factors $(\bm{z}_i,\bm{u}_i)$ and cell-level factors ($\bm{v}_{ic}, \bm{f}_{ic}$) are generated from sub-Gaussian distributions. Moreover, the diagonal entries of $\bm{\Sigma}_{\bm{x}^*}$ defined in \eqref{eq:sigma_target} below are uniformly bounded by a constant $\sigma_{x,\max}^2$, and the coefficients satisfy $\twonorm{\bm{\beta}_{anc(j)}}\vee\twonorm{\bm{\beta}_m}\le C_\beta$ for all $j\in[p]$ and $\twonorm{(\bm{I}-\bm{B})^{-1}}\le C_B$.
\end{assumption}

With the assumption, we analyze the bias induced by EIV in both SEM and outcome regressions. First, let $\tilde{\bm{x}}^*_i$ denote the $i$th row of $\tilde{\bm{X}}^*=\bm{Q}^*\bm{X}^*$. We also define
\begin{equation}\label{eq:sigma_target}
\bm{\Sigma}_{\bm{x}^*}\coloneqq\frac{1}{N}\sum_{i=1}^N\mathbb{E}\left[
\tilde{\bm{x}}^*_i\tilde{\bm{x}}^{*\top}_i
\right]    
\end{equation}
and $\bm{\Sigma}_{\bm{x}_{anc(j)}^*}$ as the ancestor node sub-matrix of $\bm{\Sigma}_{\bm{x}^*}$. We then have the following results specifying probability limits of the sample moments of the corrupted variable clarifying the EIV bias and motivating the bias correction in Section~\ref{sec:eiv_correction}.


\begin{prop}[Asymptotic EIV bias]\label{prop:eiv_bias}
Under Assumption~\ref{assump:noise}--\ref{assump:subgaussian} and  $p/N=o(1)$:\\
\textbf{(i)}. For each $j$, define $\bar{\bm{\Sigma}}_{\delta,anc(j)} = \frac{1}{N}\sum_{i=1}^N \frac{1}{n_i}\bm{\Sigma}_{\delta_i,anc(j)}$
and let $\bm{\Sigma}_{\delta_i,anc(j)}$ denote the submatrix of $\bm{\Sigma}_{\delta_i}$ indexed by
$anc(j)$. Then, we have 
\begin{align}
\twonorm{\frac{1}{N} \bar{\bm{X}}_{anc(j)}^\top \hat{\bm{Q}} \bar{\bm{X}}_{anc(j)} -\left(\bm{\Sigma}_{\bm{x}_{anc(j)}^*} + \bar{\bm{\Sigma}}_{\delta,anc(j)}\right)}
    &=o_p(1)\label{eq:denom_bias}\\
\twonorm{\frac{1}{N} \bar{\bm{X}}_{anc(j)}^\top \hat{\bm{Q}} \bar{\bm{x}}_j-\bm{\Sigma}_{\bm{x}_{anc(j)}^*} \bm{\beta}_{anc(j)}}
     &=o_p(1) \label{eq:num_unbiased}
\end{align}

\textbf{(ii)}. Define $\bar{\bm{\Sigma}}_\delta = \frac{1}{N}\sum_{i=1}^N \frac{1}{n_i}\bm{\Sigma}_{\delta_i}$, we have
\begin{align}
\twonorm{\frac{1}{N} \bar{\bm{X}}^\top \hat{\bm{Q}} \bar{\bm{X}}-\left(\bm{\Sigma}_{\bm{x}^*} + \bar{\bm{\Sigma}}_\delta \right)}
     &=o_p(1)\label{eq:denom_bias_outcome}\\
\twonorm{\frac{1}{N} \bar{\bm{X}}^\top \hat{\bm{Q}} \bm{y}-\bm{\Sigma}_{\bm{x}^*} \bm{\beta}_m}
     &=o_p(1)\label{eq:num_unbiased_outcome}
\end{align}

\end{prop}

Thus, even after adjusting for the latent confounder, naive regression approach for \eqref{eq:shifted_sem_fwl} and \eqref{eq:shifted_outcome_fwl} remains inconsistent because the predictor Gram matrix contains an additive technical-error covariance  $\bar{\bm{\Sigma}}_{\delta,anc(j)}$ (or $\bar{\bm{\Sigma}}_\delta$). This motivates subtracting an estimate of the technical-error covariance from the empirical Gram matrix.

\ignore{
The results above show that estimation based on the observed regressors leads to biased estimation. Notably, the naive OLS estimators for the SEM and outcome regressions are biased even under regimes where $p/N=o(1)$. Specifically, the estimators concentrate around an attenuated target rather than the true parameter, satisfying:
\begin{align*}
\left\| \hat{\bm{\beta}}_{anc(j),OLS} - \left\{\bm{\Sigma}_{\bm{x}_{anc(j)}^*} + \bar{\bm{\Sigma}}_{\delta,anc(j)} \right\}^{-1} \bm{\Sigma}_{\bm{x}_{anc(j)}^*} \bm{\beta}_{anc(j)} \right\|_2 &= o_p(1), \\
\left\| \hat{\bm{\beta}}_{m,OLS} - \left\{\bm{\Sigma}_{\bm{x}^*} + \bar{\bm{\Sigma}}_\delta \right\}^{-1} \bm{\Sigma}_{\bm{x}^*} \bm{\beta}_m \right\|_2 &= o_p(1),
\end{align*}
where $\hat{\bm{\beta}}_{anc(j),OLS} \coloneqq
\left( \bar{\bm{X}}_{anc(j)}^\top \hat{\bm{Q}} \bar{\bm{X}}_{anc(j)} \right)^{-1} \bar{\bm{X}}_{anc(j)}^\top \hat{\bm{Q}} \bar{\bm{x}}_j$ and $\hat{\bm{\beta}}_{m,OLS} \coloneqq
    \left( \bar{\bm{X}}^\top \hat{\bm{Q}} \bar{\bm{X}} \right)^{-1} \bar{\bm{X}}^\top \hat{\bm{Q}} \bm{y}$.

}

\subsection{EIV-corrected sparse SEM and outcome regression}\label{sec:eiv_correction}
We now provide a method to correct for the EIV biases, following a similar strategy to
\cite{loh2011high,datta2017cocolasso}. 

The SEM and outcome regressions in \eqref{eq:shifted_sem_fwl} and \eqref{eq:shifted_outcome_fwl}  may be high-dimensional: the number \(q_j\) of candidate ancestral predictors in the local SEM and the total number \(p\) of genes in the outcome regression may exceed $N$. Had we had access to the oracle regressor ($\bm{Q}^* \bm{X}^*_{anc(j)}$) in the SEM regression, we could have formulated the LASSO for the structural parameter as
\begin{equation}
    \min_{\bm{\beta}_{anc(j)}} \left\{ \frac{1}{2} \bm{\beta}_{anc(j)}^\top \bm{S}^*_j \bm{\beta}_{anc(j)} - \bm{\rho}_{\beta}^\top \bm{\beta}_{anc(j)} + \lambda_j \|\bm{\beta}_{anc(j)}\|_1  \right\},
\end{equation}
where $\bm{S}^*_j = \frac{1}{N} \bm{X}_{anc(j)}^{*\top} \bm{Q}^* \bm{X}_{anc(j)}^*$ and $\bm{\rho}_{\beta} = \frac{1}{N} \bm{X}_{anc(j)}^{*\top} \bm{Q}^* \bar{\bm{x}}_j$ are the oracle sample moments, whose probability limits are $\bm{\Sigma}_{\bm{x}^*_{anc(j)}}$ and $\bm{\Sigma}_{\bm{x}^*_{anc(j)}}\bm{\beta}_{anc(j)}$ respectively.

Likewise, for the outcome regression, the LASSO can be formulated as 
\begin{equation}
    \min_{\bm{\beta}_m} \left\{ \frac{1}{2} \bm{\beta}_m^\top \bm{S}^* \bm{\beta}_m - \bm{\rho}_{\beta_m}^\top \bm{\beta}_m + \lambda \|\bm{\beta}_m\|_1  \right\},
\end{equation}
where $\bm{S}^* = \frac{1}{N} \bm{X}^{*\top} \bm{Q}^* \bm{X}^*$ and $\bm{\rho}_{\beta_m} = \frac{1}{N} \bm{X}^{*\top} \bm{Q}^* \bm{y}$.

However, we only have access to the corrupted regressors, as shown in \eqref{eq:SEM_error} and \eqref{eq:outcome_error}, leading to EIV bias, as demonstrated in Proposition  \ref{prop:eiv_bias}. 
We now correct for such bias motivated by the results in Proposition  \ref{prop:eiv_bias}  and the previous EIV regression literature including \cite{loh2011high,datta2017cocolasso}. 
Unlike standard EIV methods that treat the measurement-error covariance as known, the replicated design allows us to estimate \(\Sigma_{\delta_i}\). The key is to first remove latent within-subject heterogeneity using the latent cell-level factor recovered from cell-scale SVA introduced in Section~\ref{sec:cell_sva}.

For the outcome  regression \eqref{eq:shifted_outcome_fwl}, we correct for the bias by employing: 
\begin{equation}
    \hat{\bm{S}} = \frac{1}{N} \bar{\bm{X}}^\top \hat{\bm{Q}} \bar{\bm{X}} - \frac{1}{N} \sum_{i=1}^N \frac{1}{n_i}\hat{\bm{\Sigma}}_{\delta_i}
\end{equation}
where $\hat{\bm{\Sigma}}_{\delta_i}$ is the residual covariance matrix regressing out $\bm{V}_i$ and $\hat{\bm{F}}_i$ from $\bm{X}^{(i)}\in \mathbb{R}^{n_i\times p}$. Specifically, we calculate the pure technical error variance as:
\begin{align}\label{eq:sub_cov}
    \hat{\bm{\Sigma}}_{\delta_i} &= \text{diag}\left( \frac{1}{n_i-l-r-1} \bm{X}^{(i)\top} \left( \bm{I}_{n_i} - \bm{P}_{[\bm{1}, \bm{V}_i, \hat{\bm{F}}_i]} \right) \bm{X}^{(i)} \right).
\end{align}

Because the subtraction of $\hat{\bm{\Sigma}}_{\delta_i}$ may cause $\hat{\bm{S}}$ to lose positive semi-definiteness, we apply a Convex Conditioned Lasso (CoCoLasso) projection \citep{datta2017cocolasso}:
\begin{equation}
\tilde{\bm{S}}=\arg\min_{\bm{S}\geq\0}{\norm{\hat{\bm{S}} -\bm{S}}_{\rm max}}.
\end{equation}

Then, the EIV-corrected LASSO for the outcome regression is convex and computed via:
\begin{equation}\label{eq:eiv_pheno_lasso}
    \hat{\bm{\beta}}_m^{EIV} = \arg\min_{\bm{\beta}_m} \left\{ \frac{1}{2} \bm{\beta}_m^{\top} \tilde{\bm{S}} \bm{\beta}_m - \hat{\bm{c}}^\top \bm{\beta}_m + \lambda \|\bm{\beta}_m\|_1  \right\},
\end{equation}
where $\hat{\bm{c}} = \frac{1}{N} \bar{\bm{X}}^\top \hat{\bm{Q}} \bm{y}$requires no correction by Proposition~\ref{prop:eiv_bias}(ii).

The gene $j$ specific local SEM regression \eqref{eq:shifted_sem_fwl} can be similarly conducted by employing $anc(j)$ entry sub-matrix of $\hat{\bm{S}}$, defined as $\hat{\bm{S}}_j\in \mathbb{R}^{q_j\times q_j}$, and a CoCoLASSO projection $\tilde{\bm{S}}_j=\arg\min_{\bm{S}\geq\0}{\norm{\hat{\bm{S}}_j -\bm{S}}_{\rm max}}$. Then, defining $\hat{\bm{a}}_j = \frac{1}{N} \bar{\bm{X}}_{anc(j)}^\top \hat{\bm{Q}} \bar{\bm{x}}_j$, we solve 
\begin{equation}\label{eq:eiv_dag_lasso}
    \hat{\bm{\beta}}_{anc(j)}^{EIV}= \arg\min_{\bm{\beta}_{anc(j)}} \left\{ \frac{1}{2} \bm{\beta}_{anc(j)}^{\top} \tilde{\bm{S}}_j \bm{\beta}_{anc(j)} - \hat{\bm{a}}_j^\top \bm{\beta}_{anc(j)} + \lambda_j\|\bm{\beta}_{anc(j)}\|_1 \right\},
\end{equation}
Placing each entry $\hat{\beta}_{anc(j),k}^{EIV}$ at position $(k,j)$ across all $j\in[p]$, and setting the remaining entries to zero, assembles $\hat{\bm{B}}\in\mathbb{R}^{p\times p}$. We provide the detailed EIV-correction procedure in
Algorithm~\ref{alg:eiv_correction} with details on hyperparameter selection in Section~\ref{supp-sec:implementation}.

\begin{algorithm}[!ht]
\caption{EIV-Corrected SEM and Outcome Regression}\label{alg:eiv_correction}
\textbf{Input:} $\bar{\bm{X}}$, $\bm{Z}_{aug}$, $\hat{\bm{U}}_{aug}$, $\{\hat{\bm{F}}_i\}_{i=1}^N$ from Algorithm~\ref{alg:sva_correction}; $\{\bm{X}^{(i)},\bm{V}_i\}_{i=1}^N$; ancestor sets $\{anc(j)\}_{j=1}^p$; outcome vector $\bm{y}$; tuning parameters $\{\lambda_j\}_{j=1}^p$ and $\lambda$.\\
\textbf{Output:} $\hat{\bm{B}}$ and $\hat{\bm{\beta}}_m^{EIV}$.

\underline{Step 1 (Projection).} Set $\hat{\bm{H}}=[\bm{1},\bm{Z}_{aug},\hat{\bm{U}}_{aug}]$ and $\hat{\bm{Q}}=\bm{I}_N-\hat{\bm{H}}(\hat{\bm{H}}^\top\hat{\bm{H}})^{-1}\hat{\bm{H}}^\top$.

\underline{Step 2 (Technical noise variance).} For each $i\in[N]$, compute $\hat{\bm{\Sigma}}_{\delta_i}$ as in \eqref{eq:sub_cov}.

\underline{Step 3 (SEM regression).} For each $j\in[p]$:
\begin{enumerate}
    \item[(a)] Assemble $\hat{\bm{S}}_j=\frac{1}{N}\bar{\bm{X}}_{anc(j)}^\top\hat{\bm{Q}}\bar{\bm{X}}_{anc(j)}-\frac{1}{N}\sum_{i=1}^N\frac{1}{n_i}\hat{\bm{\Sigma}}_{\delta_i,anc(j)}$ and $\hat{\bm{a}}_j=\frac{1}{N}\bar{\bm{X}}_{anc(j)}^\top\hat{\bm{Q}}\bar{\bm{x}}_j$.
    \item[(b)] Project $\tilde{\bm{S}}_j=\arg\min_{\bm{S}\succeq\bm{0}}\|\hat{\bm{S}}_j-\bm{S}\|_{\max}$ (CoCoLasso).
    \item[(c)] Solve for $\hat{\bm{\beta}}^{EIV}_{anc(j)}$ as in \eqref{eq:eiv_dag_lasso}, with $\tilde{\bm{S}}_j,\hat{\bm{a}}_j,\lambda_j$.
    \item[(d)] Set $\hat{B}_{kj}=\hat\beta^{EIV}_{anc(j),k}$ for $k\in anc(j)$ and $\hat{B}_{kj}=0$ otherwise.
\end{enumerate}

\underline{Step 4 (Outcome regression).}
\begin{enumerate}
    \item[(a)] Assemble $\hat{\bm{S}}=\frac{1}{N}\bar{\bm{X}}^\top\hat{\bm{Q}}\bar{\bm{X}}-\frac{1}{N}\sum_{i=1}^N\frac{1}{n_i}\hat{\bm{\Sigma}}_{\delta_i}$ and $\hat{\bm{c}}=\frac{1}{N}\bar{\bm{X}}^\top\hat{\bm{Q}}\bm{y}$.
    \item[(b)] Project $\tilde{\bm{S}}=\arg\min_{\bm{S}\succeq\bm{0}}\|\hat{\bm{S}}-\bm{S}\|_{\max}$.
    \item[(c)] Solve for $\hat{\bm{\beta}}_m^{EIV}$ as in \eqref{eq:eiv_pheno_lasso}, with $\tilde{\bm{S}},\hat{\bm{c}},\lambda$.
\end{enumerate}

\textbf{Return} $\hat{\bm{B}}=(\hat{B}_{kj})_{k,j\in[p]}$ and $\hat{\bm\beta}_m^{EIV}$.
\end{algorithm}

\subsection{Estimation error of the EIV-corrected estimators}

We now establish error bounds for the estimators
$\hat{\bm{\beta}}^{EIV}_{anc(j)}$ and $\hat{\bm{\beta}}^{EIV}_m$.  Throughout this section we
 work in the high-dimensional regime, allowing $p\gg N$. For the SEM regression write $s_j\coloneqq \norm{\bm{\beta}_{anc(j)}}_0=|pa(j)|$, $B_j\coloneqq \norm{\bm{\beta}_{anc(j)}}_{\rm max}$, $\kappa_j\coloneqq 1+s_jB_j$ for $\forall j\in [p]$
and for the outcome regression $s_m\coloneqq \norm{\bm{\beta}_m}_0$, $b_m\coloneqq \norm{\bm{\beta}_m}_{\rm max}$, $\kappa_m\coloneqq 1+s_mb_m$.
We also set $s_{\max}\coloneqq \max\{s_m,s_B\}$, $s_B\coloneqq \max_{j\in[p]}s_j$ , $n_{\min}\coloneqq\min_{i}n_i$
and $d_c\coloneqq 1+l+r$. 

We also impose additional regularity assumptions for establishing the asymptotic properties of our proposed estimators. Assumption~\ref{assump:coco}, stated below, corresponds to the compatibility assumption proposed in \cite{datta2017cocolasso}, combined with sparsity and boundedness assumptions. Assumption~\ref{assump:sva_rate} specifies the conditions for the two-scale SVA in \eqref{eq:subject_recovery_hyp} and \eqref{eq:cell_recovery_hyp}.

\begin{assumption}[Design regularity]\label{assump:coco}
Assume
\begin{itemize}
\vspace{-0.3em}
\item [(i)] (Compatibility): Define $\phi_j \coloneqq \min\left\{ \frac{\bm{x}^\top\bm{\Sigma}_{\bm{x}^*_{anc(j)}}\bm{x}}{\twonorm{\bm{x}}^2}
     :  \bm{x}\in\mathbb{R}^{q_j}\setminus\{\bm{0}\},\ \frac{\onenorm{\bm{x}_{pa(j)^c}}}{\onenorm{\bm{x}_{pa(j)}}}\le 3\right\}>0$ for each $j\in[p]$ and define $\phi_m>0$ analogously with $\bm{\Sigma}_{\bm{x}^*}$ and $\mathrm{supp}(\bm{\beta}_m)$.
\item [(ii)] (Sparsity): $\max_j B_j\vee b_m=\mathcal{O}(1)$ and $s_{\max}^2\log p/N\rightarrow 0$.
\item [(iii)] (Sub-Gaussian tail bounds): With $\bm{E}^B=\bm{E}(\bm{I}-\bm{B})^{-1}$,
 $\sup_{i\in[N]}\ \sup_{\bm u\in\mathbb{S}^{p-1}}\norm{\langle\bm u,\bm\varepsilon_i^B\rangle}_{\psi_2}\le \sigma_\varepsilon<\infty$ and
$\norm{\varepsilon_{yi}}_{\psi_2}\le K_y<\infty.$   
\item [(iv)] (Bounded loadings): $\max_{k\in[p]}\twonorm{\bm{\alpha}_k}\vee\twonorm{\bm{\rho}_k}\le C_A<\infty$, and, writing $\bar{\bm{X}}=\bm{H}\bm{\Xi}+\bm{C}$ with $\bm{H}=[\bm{1},\bm{Z}_{aug},\bm{U}_{aug}]$ by Lemma~\ref{supp-lem:xbar_decomp}, $\max_{k\in[p]}\twonorm{\bm{\xi}_k}\le C_\Xi<\infty$.
\end{itemize}
\end{assumption}

\begin{assumption}[Rates for the two surrogate variable analyses]\label{assump:sva_rate}
Assume
\begin{itemize}
\vspace{-0.3em}
\item [(i)] (Subject-scale signal): For a large enough constant $c_S>0$,
\begin{equation}\label{eq:subject_signal}
    \sigma_{m+r}(\bm{\Phi}) \ge  c_S\big(\sqrt{N}+\sqrt{p}\big)\max\left\{
    1,\sqrt{N/\log p}
    \right\}.
\end{equation}
\item [(ii)] (Cell-scale signal): For a large enough constant $C_F> 0$,
\begin{equation}\label{eq:cell_signal}
    \sigma_r(\bm{\Phi}_i) \ge C_F\big(\sqrt{n_i}+\sqrt{p}\big)
    \max\left\{1,\ \frac{1}{n_{\min}}\sqrt{\frac{N}{\log p}}\right\}
    \qquad\text{for every } i\in[N].
\end{equation}
\item [(iii)] (Scaling): $\log(Np)=o(n_{\min}\wedge p)$, and $n_{\min}\gtrsim(N/\log p)^{1/3}$.
\end{itemize}
\end{assumption}



With the additional assumptions above, we can derive key theorems establishing the following high-probability bound results for our proposed estimators.

\begin{thm}[Estimation error of the EIV-corrected DAG estimator]\label{thm:coco_rate}
Let Assumptions~\ref{assump:noise}-\ref{assump:sva_rate} hold, and define $\sigma_{n_{\min}}\coloneqq \sigma_\varepsilon+\sigma_{\max}/\sqrt{n_{\min}}$. Fix a sufficiently large constant $C_\lambda>0$. Then, for every $\alpha>0$, there exist $C,c>0$ such that, with probability at least $1-Cp^{-\alpha}-Ce^{-c(N\wedge p)}$, the following holds. For every collection of tuning parameters $\{\lambda_j\}_{j\in[p]}$ satisfying $\lambda_j \ge C_\lambda \sigma_{n_{\min}}^2 \kappa_j\sqrt{\frac{\log p}{N}}$, write $\lambda_{\max}\coloneqq\max_{j\in[p]}\lambda_j$ and $\phi_{\rm min}\coloneqq\min_{j\in[p]}\phi_j$. Then
\begin{equation*}
    \max_{j\in[p]}\twonorm{\hat{\bm{\beta}}^{EIV}_{anc(j)}-\bm{\beta}_{anc(j)}}
     \le \frac{C\lambda_{\max}\sqrt{s_B}}{\phi_{\rm min}},
    \qquad
    \max_{j\in[p]}\onenorm{\hat{\bm{\beta}}^{EIV}_{anc(j)}-\bm{\beta}_{anc(j)}}
     \le \frac{C\lambda_{\max} s_B}{\phi_{\rm min}} .
\end{equation*}
\end{thm}

\begin{thm}[Estimation error of the EIV-corrected outcome estimator]\label{thm:coco_rate_outcome}
Under the conditions of Theorem~\ref{thm:coco_rate}, with $\sigma_{n_{\min}}^\dagger\coloneqq
\sigma_{n_{\min}}+K_y$, the following holds on the same event as Theorem~\ref{thm:coco_rate}: for
every $\lambda \ge C_\lambda (\sigma_{n_{\min}}^{\dagger})^2 \kappa_m\sqrt{\frac{\log p}{N}},$
\begin{equation*}
    \twonorm{\hat{\bm{\beta}}^{EIV}_{m}-\bm{\beta}_{m}} \le \frac{C\lambda\sqrt{s_m}}{\phi_m},
    \qquad
    \norm{\hat{\bm{\beta}}^{EIV}_{m}-\bm{\beta}_{m}}_1 \le \frac{C\lambda s_m}{\phi_m} .
\end{equation*}
\end{thm}
The established estimation errors match those provided in \cite{datta2017cocolasso}. The finite-sample error bounds in the theorems hold under the scaling $s_{\max}^2 \log p\ll N$, which permits $p$ to grow exponentially in $N$. This also aligns with our motivating example's setup, in that we expect the number of direct parent genes ($s_j$) in each local SEM regression, and the number of directly influencing genes ($s_m$) for the subject-level outcome in the outcome regression, to be relatively small compared to the total number of genes ($p$) analyzed.


For the case when high levels of correlation across variables are of concern, we additionally provide an Elastic-net \citep{zou2005regularization} extension of our method and establish estimation error results for the extensions of \eqref{eq:eiv_dag_lasso} and \eqref{eq:eiv_pheno_lasso} in Section~\ref{supp-sec:enet}.


We now provide the following key convergence results on the estimation of path-specific and total causal effects over the augmented graph $\mathcal{G}^+$.

\begin{cor}[Consistency of path-specific causal effects]\label{cor:pce_rate}
Fix a path $\pi=(j_1,j_2,\dots,j_k,y)\in\Pi_{j_1\to y}$, let
$\widehat{\mathrm{PCE}}_\pi\coloneqq\big(\prod_{l=1}^{k-1}\hat B_{j_l,j_{l+1}}\big)\hat\beta_{m,j_k}$. Under
Assumptions~\ref{assump:noise}--\ref{assump:sva_rate}, on the same
event as Theorems~\ref{thm:coco_rate}, with
$\beta\coloneqq \max_jB_j\vee b_m$ and $\epsilon\coloneqq
\big(C\lambda_{\max}\sqrt{s_B}/\phi_{\min}\big)\vee\big(C\lambda\sqrt{s_m}/\phi_m\big)$,
\begin{equation*}
    \big|\widehat{\mathrm{PCE}}_\pi-\mathrm{PCE}_\pi\big|\ \le\ k(2\beta)^{k-1} \epsilon .
\end{equation*}
\end{cor}

For the Corollary on TCE below, define $\gamma_j\coloneqq1-\norm{\bm{G}_j}_{\infty}$ and $\gamma_m\coloneqq1-\norm{\bm{G}_m}_{\infty}$ where $\bm{G}_j=\mathrm{Cov}\big(\bm{x}^*_{S_j^c},\bm{x}^*_{S_j}\big)\big(\bm{\Sigma}_{\bm{x}^*_{S_j}}\big)^{-1}$, $\bm{G}_j=\mathrm{Cov}\big(\bm{x}^*_{S_m^c},\bm{x}^*_{S_m}\big)\big(\bm{\Sigma}_{\bm{x}^*_{S_m}}\big)^{-1}$, and $S_j\coloneqq pa(j)$, $S_j^c\coloneqq anc(j)\setminus pa(j)$ and $S_m\coloneqq \mathrm{supp}(\bm{\beta}_m)$, $S_j^c\coloneqq [p]\setminus \mathrm{supp}(\bm{\beta}_m)$.

\begin{cor}[Consistency of total causal effects]\label{cor:tce_rate}
Under Assumptions~\ref{assump:noise}--\ref{assump:sva_rate} and \ref{supp-assump:irrep}, on the event of
Theorem~\ref{thm:coco_rate}, for every $\{\lambda_j\}_{j\in[p]}$ and $\lambda$ satisfying 
\begin{equation*}
    \lambda_j\ \ge\ \frac{C_\lambda'}{\gamma_{\min}} \sigma_{n_{\min}}^2 \kappa_j\sqrt{\frac{\log p}{N}}
    \quad\forall j\in[p],
    \qquad
    \lambda\ \ge\ \frac{C_\lambda'}{\gamma_{\min}} (\sigma_{n_{\min}}^\dagger)^2 \kappa_m\sqrt{\frac{\log p}{N}},
\end{equation*}
where $\gamma_{\min}>0$ satisfies $\min_{j\in[p]}\gamma_j\wedge\gamma_m\geq \gamma_{\min}$, writing $k_{j,\max}$ for the length of the longest path in $\Pi_{j\to y}$ and $\beta,\epsilon$ as in
Corollary~\ref{cor:pce_rate}, simultaneously for every  $j\in[p]$, we have
\begin{equation*}
    \big|\widehat{\mathrm{TCE}}_j-\mathrm{TCE}_j\big|\ \le\ \big|\Pi_{j\to y}\big| k_{j,\max} (2\beta)^{k_{j,\max}-1} \epsilon.
\end{equation*}
\end{cor}
Notice that the errors of PCE and TCE are tied to the estimation errors of the coefficients $\bm{B}$ and $\bm{\beta}_m$, as controlled by Theorems~\ref{thm:coco_rate} and \ref{thm:coco_rate_outcome}. The PCE error bound carries a multiplier $k$, the length of the path $\pi$, while the TCE error bound additionally scales with the number of paths $\big|\Pi_{j\to y}\big|$ and their maximum length $k_{j,\max}$. We assume that neither quantity grows as $n,p\to\infty$, which is reasonable since gene regulatory networks typically do not exhibit long directed cascades \citep{alon2007network}, and the number of regulatory genes reaching a given gene remains bounded even as network size grows \citep{aguirre2026regulatory}.

 \section{Simulation studies}\label{sec:simul}

\subsection{Simulation setup}

We compare the proposed SVA+EIV estimator against three ablations -- Naive (no confounder or measurement-error correction), SVA (confounder correction only), and EIV (measurement-error correction only) -- across a range of sample sizes. 

We fix a DAG on $p=60$ genes. For $j=2,\dots,p$, the parent count $n_{pa}(j)\stackrel{iid}{\sim}\mathrm{Unif}\{0,1,2\}$ and the parents are drawn uniformly without replacement from $\{1,\dots,j-1\}$, with edge weights $B_{kj}=\pm1$, $\Pr(B_{kj}=-1)=0.7$. Subject-level states follow $\bm{M}=(\bm{1}\bm{\mu}^\top+\bm{Z}\bm{\Gamma}+\bm{U}\bm{\Lambda}+\bm{E})(\bm{I}-\bm{B})^{-1}$, with $k=m=1$: $z_i,u_i\stackrel{iid}{\sim}N(0,1)$, $\mu_j,\gamma_j\stackrel{iid}{\sim}N(0,0.5^2)$, $|\lambda_j|\stackrel{iid}{\sim}U(1.5,2)$ with random sign, and $\varepsilon_{ij}\stackrel{iid}{\sim}N(0,0.5^2)$. See \textbf{Fig.}~\ref{supp-fig:simul_result_90},\ref{supp-fig:simul_result_120} for larger $p$ results ($p=90,120$). Each subject has $n_i\stackrel{iid}{\sim}\mathrm{Unif}\{20,\dots,150\}$ cells, following the measurement error model, $x_{ijc}=m_{ij}+\bm{\rho}_j^\top\bm{v}_{ic}+\bm{\alpha}_j^\top\bm{f}_{ic}+\delta_{ijc}$, with $l=r=1$: $v_{ic}\stackrel{iid}{\sim}N(0,1)$ and $f_{ic}=\nu_i+\eta_{ic}$ with subject-specific mean $\nu_i\stackrel{iid}{\sim}N(0,0.5^2)$ and within-subject noise $\eta_{ic}\stackrel{iid}{\sim}N(0,1)$; $\rho_j\stackrel{iid}{\sim}U(1.5,2)$, $|\alpha_j|\stackrel{iid}{\sim}U(1.5,2)$ with random sign, and $\delta_{ijc}\sim N(0,\sigma^2_{\delta_{ij}})$ with $\sigma^2_{\delta_{ij}}\stackrel{iid}{\sim}U(4,8)$. Following $y_i=\bm{m}_i^\top\beta_m+z_i\beta_z+u_i\beta_u+\varepsilon_{yi}$, where $\beta_m$ has $5$ nonzero entries set to $\pm1.5$ with random sign, $\beta_z\sim U(1.5,2)$, $|\beta_u|\sim U(1.5,2)$ with random sign, and $\varepsilon_{yi}\stackrel{iid}{\sim}N(0,0.5^2)$. We vary $N\in\{40,120,360\}$ with $p=90$, $\bm{B}$, and $(\beta_m,\beta_z,\beta_u)$ fixed, generating $200$ simulation runs per $N$. All four estimators share an elastic-net penalty ($\alpha_{\mathrm{elastic}}=0.5$) with tuning parameters $\lambda_j,\lambda$ selected based on $5$-fold cross-validation (see Section~\ref{supp-sec:implementation} for details). The four methods all employ the true ancestor structure of the generated graph. We evaluate $\twonorm{\sin\Theta(\tilde{\bm{U}}_{aug},\hat{\bm{U}}_{aug})}$ and $\twonorm{\sin\Theta(\tilde{\bm{F}}_i,\hat{\bm{F}}_i)}$, $\Fnorm{\hat{\bm{B}}-\bm{B}}$, $\twonorm{\hat\beta_m-\beta_m}$ and $\twonorm{\widehat{\mathrm{TCE}}-\mathrm{TCE}}$. All the metrics are averaged over the $200$ runs at each setting.

\subsection{Confounder recovery and estimation error comparison}

\begin{figure}[!ht]\centering
   \includegraphics[width=1\textwidth]{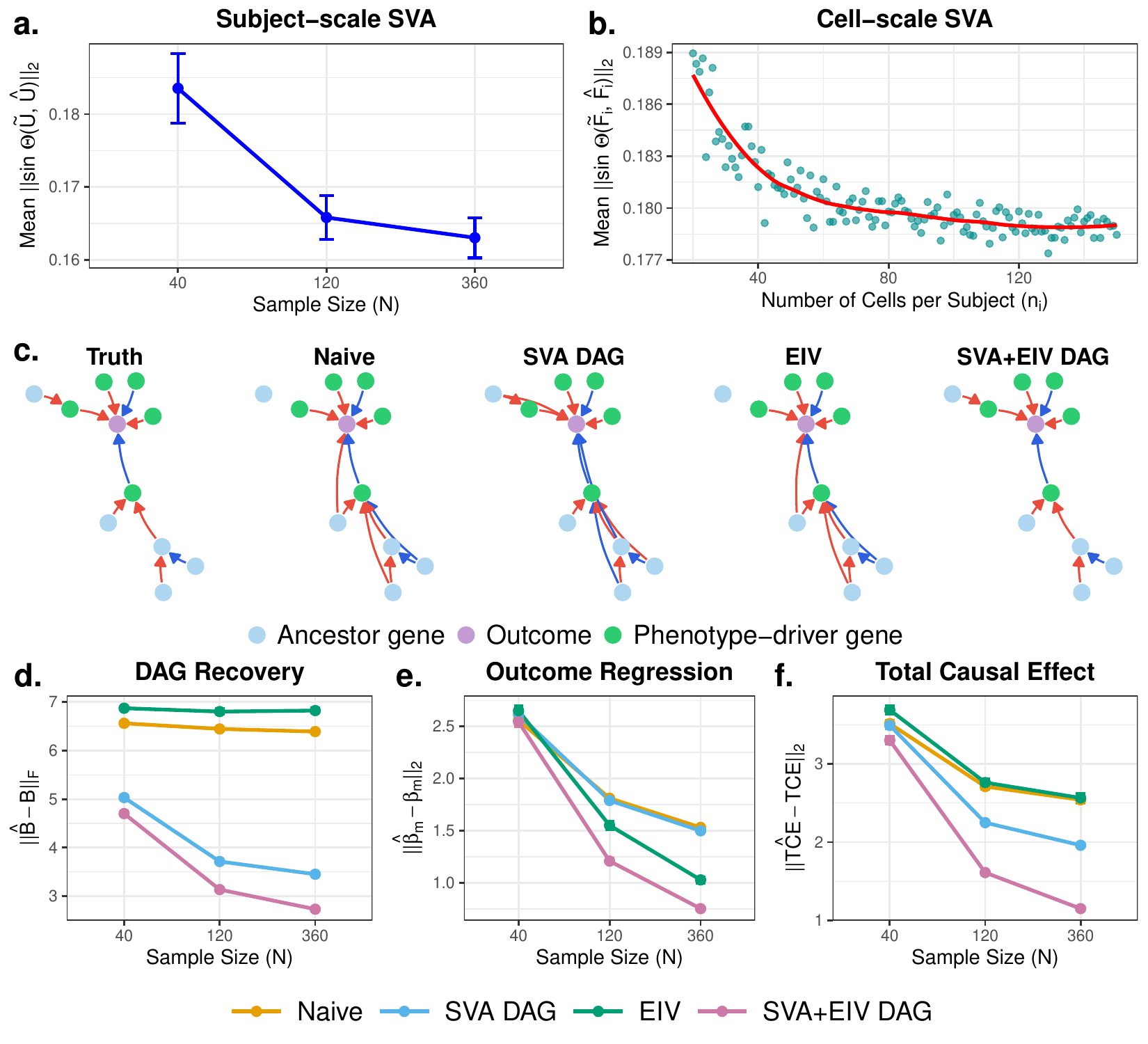}
\caption{\textbf{Main simulation results with gene number $p=60$.} \textbf{a.} Summary of $\twonorm{\sin\Theta(\tilde{\bm{U}}_{aug},\hat{\bm{U}}_{aug})}$ (y-axis) across sample size $N$ throughout 200 replicates. \textbf{b.} Summary of $\twonorm{\sin\Theta(\tilde{\bm{F}}_i,\hat{\bm{F}}_i)}$ (y-axis) across cell number $n_i$. Each dot represents the mean error across all subjects (and simulation replications) sharing the same cell count $n_i$ and red line denotes the LOESS fit. \textbf{c.} Illustration of the true and learned graphs across all four methods. The outcome variable ($y$, colored purple), its immediate parent genes (green), and their ancestors up to two hops back (light blue) are displayed. Edge colors denote the sign of the coefficient (red: negative, blue: positive). \textbf{d,e,f.} Summary of $\Fnorm{\hat{\bm{B}}-\bm{B}}$, $\twonorm{\hat\beta_m-\beta_m}$, and $\twonorm{\widehat{\mathrm{TCE}}-\mathrm{TCE}}$ (y-axis) across sample size $N$ (x-axis) and the four methods (colors), respectively.
} \label{fig:simul_result}
\end{figure}

We first checked whether the unmeasured confounders can be recovered through our proposed SVA approach. It turned out that $\twonorm{\sin\Theta(\tilde{\bm{U}}_{aug},\hat{\bm{U}}_{aug})}$ and $\twonorm{\sin\Theta(\tilde{\bm{F}}_i,\hat{\bm{F}}_i)}$ decrease toward zero as 
$N$ amd $n_i$ increases, respectively (\textbf{Fig.}~\ref{fig:simul_result}a,b), corroborating Theorems~\ref{thm:u_recovery} and \ref{thm:cell_recovery}.
We next evaluated how well the graph $\mathcal{G}^+$ can be recovered across methods. A visual comparison for $N=40$ and $p=60$, involving the outcome, its immediate influencing genes, and their parent and grandparent genes, shows that our proposed method correctly recovers the edges (\textbf{Fig.}~\ref{fig:simul_result}c). A systematic comparison across simulation runs and settings shows that our proposed method achieves uniformly smaller DAG recovery error, outcome regression error, and TCE error (\textbf{Fig.}~\ref{fig:simul_result}d-f, \textbf{Fig.}~\ref{supp-fig:simul_result_90},\ref{supp-fig:simul_result_120}). Notably, the DAG and TCE estimation errors of Naive and EIV show slower decrease than those of SVA and SVA+EIV as $N$ increases, reflecting the influence of unmeasured confounding. The gap between SVA and our method persists as $N$ increases, indicating remaining EIV bias. A more detailed comparison of edge-specific bias in the SEM and outcome regressions similarly shows that the baseline approaches yield biased edge coefficients that do not shrink as $N$ increases, while our approach shows reduced bias across all settings (\textbf{Fig.}~\ref{supp-fig:simul_result_bias_60}--\ref{supp-fig:simul_result_bias_120}). An additional sensitivity analysis examines the effect of ancestor-set misspecification on SVA+EIV, by adding false ancestor nodes to, or omitting true ancestor nodes from, every gene's ancestors. Even when 30\% of the nodes are perturbed in this way, SVA+EIV continues to achieve lower DAG and TCE errors than the Naive or EIV methods and perturbing 10\% (20\%) of the nodes achieve comparable DAG (TCE)  error to SVA approach (\textbf{Fig.}~\ref{supp-fig:appl_ancestor_mis}).

\section{Application to Population-Scale scRNA-seq Data from an Acute Myeloid Leukemia Cohort}

\subsection{Description of the data}

We applied the proposed method to population-scale scRNA-seq data from \cite{van2019single} (2019) to characterize regulatory pathways linking gene expression to blast count, a clinically relevant measure of leukemic burden. The processed data involves $N=30$ samples from bone marrow tissues of AML disease cohort with scRNA-seq data recorded for each sample. Specifically, there are $p=25$ genes and $n_i\in [127,2256]$ number of cells for each sample $i\in [30]$ and the Unique Molecular Identifier (UMI) counts were measured for each cell and gene of a sample. Each sample has a blast count measurement $y_i\in [0,1]$ as an outcome, and there are covariates in both scales: cell cycle and cell type labels for the cell-level covariate $(\bm{v}_{ic})$ and age and gender for the sample-level covariate ($\bm{z}_{i}$). For the gene expression levels, a naive log transformation of UMI counts can incur artificial variation in expression levels \citep{townes2019feature}. Therefore, we obtained normalized expression level $x_{ijc}$ from a PLN fit \citep{chiquet2019variational} on the UMI count denoted as $C_{ijc}$ drawn from $C_{ijc}|x_{ijc},l_{ic}\stackrel{i.i.d}{\sim}Pois\left\{l_{ic}\exp{(x_{ijc})}\right\},$ where $l_{ic}$ corresponds to the library size of cell $c$ for subject $i$. For details on data processing we refer to Section~\ref{supp-sec:processing_app}.

For the external information on the topological structure of the DAG over genes, we employ the ancestral structure of genes that can be learned from Perturb-seq data of K562 cells \citep{replogle2022mapping}. We specifically applied a confounder-robust causal gene network recovery method ARGEN by \cite{park2026causal} to the perturbational data and retained the ancestor gene sets of each gene: $\{j\in [25]|anc(j)\subset [p]\}$. Here, K562 was specifically employed, because it is a well-established multipotent hematopoietic progenitor model capable of both erythroid and myeloid differentiation \citep{andersson1979induction,tabilio1983myeloid}, matching the cell populations that we leveraged for our population-scale data.

\subsection{Path-Specific Effects of Genes on Blast Count}

\begin{figure}[!htbp]\centering
   \includegraphics[width=1.0\textwidth]{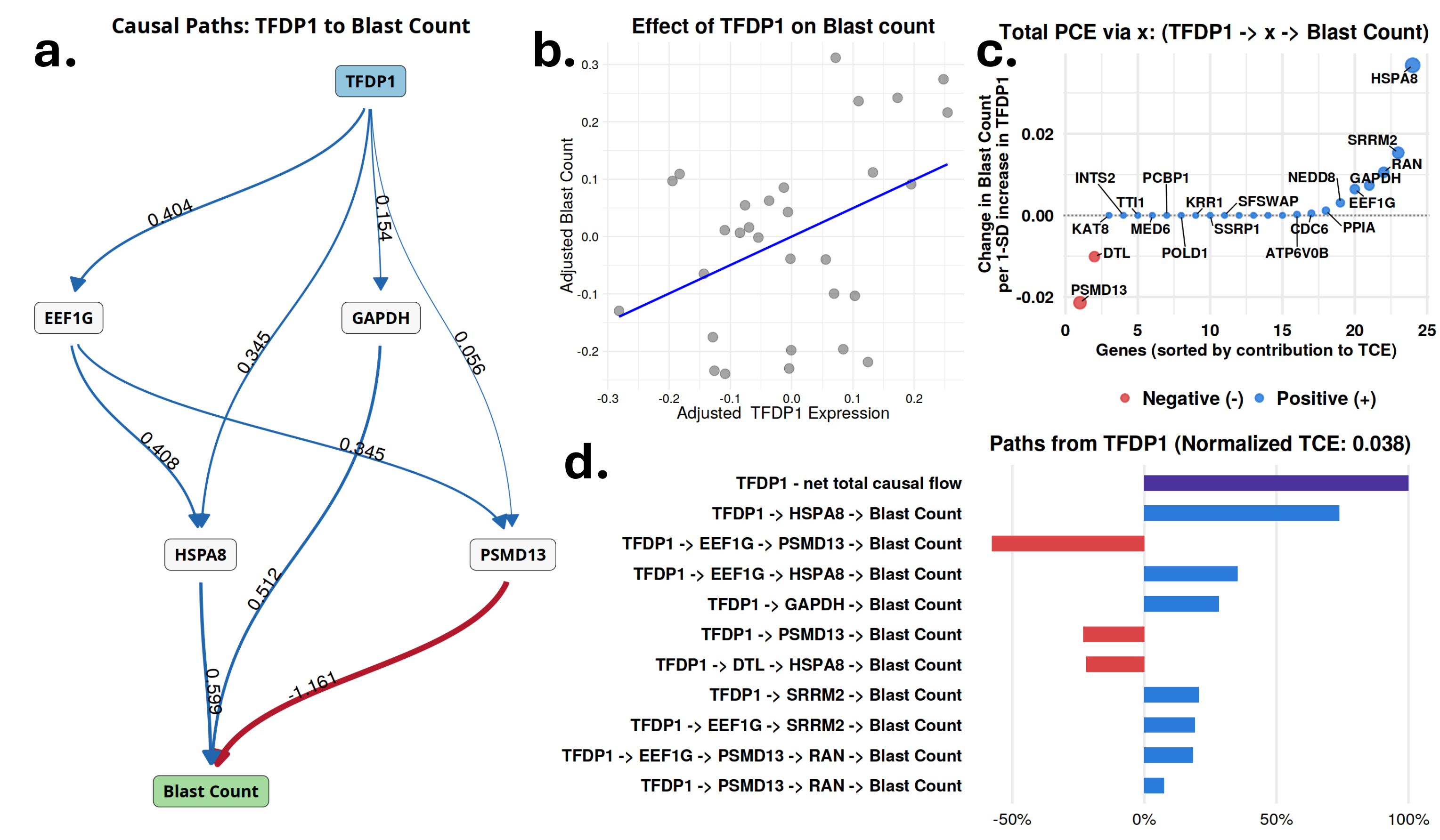}
\caption{\textbf{Path analysis results on AML cohort scRNA-seq data.}\textbf{a.} Subgraph of $\hat{\mathcal{G}}^+$ showing the top 5 paths from TFDP1 to blast count ($y$). \textbf{b.} Scatter plot of adjusted blast count (y-axis) against TFDP1 expression level (x-axis). \textbf{c.} Total PCE from TFDP1 to blast count mediated through a given gene $x$. Each dot represents the total PCE mediated through a single gene $x$. \textbf{d.} Decomposition of TFDP1's TCE (top bar) into its constituent PCEs, with only the top 10 paths by PCE magnitude shown.
} \label{fig:app_result}
\end{figure}
We applied our proposed method to the clinical-scale scRNA-seq data of \cite{van2019single}. After the data processing step detailed in Section~\ref{supp-sec:processing_app}, we focused on three transcription factor (TF) transcribing genes (TFDP1, PCBP1, KAT8) and their target genes. Target genes were defined as those showing enriched ChIP-seq signal of the TFs on their promoter regions based on ENCODE \citep{encode2012integrated} and changing significantly upon knockdown of the TF genes in Perturb-seq data from the K562 cell line \citep{replogle2022mapping}. For details on the implementation of the method and gene selection, see Section~\ref{supp-sec:processing_app}.

The estimated DAG $\hat{\bm{B}}$ over $p=25$ genes contained 103 nonzero edges with 4.12 parent genes on average per each gene. Also, 10 genes of the 25 showed a non-zero direct effect on blast count. Among the 21 genes with a nonzero estimated TCE, the sign match rate between TCE and marginal effect of adjusted expression on adjusted blast count was 81\%, with correlation $\rho=0.57$, while adjusted for $\hat{\bm{U}}_{aug}$ and $\bm{Z}_{aug}$ (\textbf{Fig.}~\ref{supp-fig:appl_tce_marginal}). This suggests that the path-specific effects are decomposing the total marginal effect of a gene on the outcome. A stability analysis of the estimated coefficients across 200 bootstrap replicates shows that, among edges identified as non-zero in the original fit, each was selected as non-zero in $34\%$ of replicates on average. Among
these non-zero edges, the majority sign observed across replicates in which the edge
was selected matched the sign of the original point estimate in $95\%$ of cases (\textbf{Fig.}~\ref{supp-fig:appl_bootstrap}). In addition, the singular values obtained from the two-scale SVA show clear elbow points in the scree plots (\textbf{Fig.}~\ref{supp-fig:appl_confounder_valid}a-c), indicating hidden low-rank factors. The recovered confounder $\hat{\bm{U}}_{aug,1}$ further shows strong association $\rho=0.5$ with blast count, revealing the importance of adjusting for the unmeasured factors (\textbf{Fig.}~\ref{supp-fig:appl_confounder_valid}d). 

We next focused on the pathway driven by TFDP1, which showed the largest standardized TCE magnitude on blast count among the three TFs (for results on the other two TFs, see \textbf{Fig.}~\ref{supp-fig:app_result_kat8}, \ref{supp-fig:app_result_pcbp1}). TFDP1 is a positive regulator of hematopoietic stem and progenitor cell (HSPC) proliferation \citep{tran2024vivo}, and dysregulated HSPC proliferation is a widely recognized hallmark of AML development \citep{bonnet1997human}.

The top five paths from TFDP1 to blast count with the greatest path-specific effect magnitudes (\textbf{Fig.}~\ref{fig:app_result}a) involve the downstream genes HSPA8, PSMD13, EEF1G and GAPDH.  The estimated TCE for TFDP1, $\hat{\mathrm{TCE}}_{\rm TFDP1}=0.28$, indicates a 28 percentage-point increase in blast count per unit increase in TFDP1 expression, consistent with the positive association between adjusted TFDP1 expression and adjusted blast count (\textbf{Fig.}~\ref{fig:app_result}b). Among the genes lying between TFDP1 and blast count, HSPA8, and PSMD13 showed the strongest evidence as hub nodes, as measured by the total path-specific effect through each gene (\textbf{Fig.}~\ref{fig:app_result}c), defined as the sum of all path-specific effects of TFDP1 on blast count passing through that gene. A more detailed, path-by-path analysis showed that $\{\mbox{TFDP1}\rightarrow\mbox{HSPA8}\rightarrow \mbox{Blast count}\}$ carried largest positive effect, while $\{\mbox{TFDP1}\rightarrow \mbox{EEF1G}\rightarrow \mbox{PSMD13}\rightarrow \mbox{Blast count}\}$ carried largest negative effect (\textbf{Fig.}~\ref{fig:app_result}d). In addition, the path $\{\mbox{TFDP1}\rightarrow\mbox{GAPDH}\rightarrow \mbox{Blast count}\}$ showed a separate route involving neither PSMD13 nor HSPA8 with notable positive path-specific effect.  The decomposition illustrates that a positive total effect can arise from the net contribution of multiple pathways with opposing signs.

The genes on these paths align with the established leukemia literature. HSPA8 is overexpressed in AML and independently associated with worse patient survival \citep{li2021high}. PSMD13 is suppressed in AML and required for normal blood cell development, so its loss is consistent with impaired regulation of blast growth \citep{gao2023interlukin}. GAPDH is overexpressed in drug-resistant leukemia cells, where it protects against caspase-independent cell death, consistent with a growth-promoting role along this path \citep{colell2009novel}.

Notably, HSPA8 and GAPDH, the two genes carrying positive path-specific effects on blast count, are both established druggable targets, with existing small-molecule inhibitors (VER-155008 for HSPA8; koningic acid for GAPDH) shown to suppress their downstream oncogenic activity \citep{wu2024hspa8,liberti2017predictive}. This suggests the $\{\mbox{TFDP1}\rightarrow\mbox{HSPA8}\rightarrow \mbox{Blast count}\}$ and $\{\mbox{TFDP1}\rightarrow\mbox{GAPDH}\rightarrow \mbox{Blast count}\}$ paths as pharmacologically tractable routes to reduce TFDP1's downstream effect on blast count. TFDP1 itself would be a harder and less precise target: it lacks a well-defined small-molecule binding pocket, and intervening on it would simultaneously perturb both of these growth-promoting paths and the opposing, growth-suppressing $\{\mbox{TFDP1}\rightarrow \mbox{EEF1G}\rightarrow\mbox{PSMD13}\rightarrow \mbox{Blast count}\}$ path, effects a single regression coefficient could not distinguish.


\section{Discussion}

We developed a framework for integrating external perturbational information with
population-scale single-cell data for causal path analysis in the presence of multiscale
unmeasured confounding and measurement error. Rather than assuming that an entire
perturbational network transfers to the target population, we use external ancestral
relationships to constrain candidate causal directions and re-estimate the direct edges and
their effects from the population data. A two-scale surrogate-variable procedure addresses
latent heterogeneity at the subject and cell levels, while the replicated measurements
enable estimation and correction of technical measurement error. The simulations illustrate that confounding adjustment and measurement-error correction address complementary sources of bias. Correcting either one alone leaves substantial estimation error, whereas their combination improves estimation of the network and gene--outcome effects. The AML application further illustrates how the proposed framework can use external perturbational information to estimate a
target-population regulatory network and decompose gene--phenotype relationships into
distinct directed pathways. 

Several limitations should be noted. Causal interpretation relies on the structural
transportability assumption that the external ancestral scaffold contains the true
target-population parents and preserves their upstream--downstream ordering. The current
framework also assumes linear structural and outcome models. Causal interpretation also requires that the clinical outcome is downstream of the gene-expression network. This assumption may be particularly consequential for cross-sectional disease phenotypes.

Future work could relax the structural transportability assumption by allowing uncertainty in the external ancestral scaffold, extend the framework to nonlinear models and
more general outcome types, and develop direct uncertainty quantification for path-specific effects. 

\section*{Supplementary Materials} The Supplementary Materials include extensions to Elastic-net, all the proofs, additional simulation results, details on data processing and implementation, and additional figures.

\bibliography{bibtex}

\end{document}